\documentclass[pdflatex,sn-mathphys,Numbered]{sn-jnl}% Math and Physical Sciences Reference Style

\usepackage{xcolor}
\usepackage{manyfoot}
\usepackage{newtxtext,newtxmath}
\usepackage{subcaption}
\usepackage{bookmark}
\usepackage[noabbrev]{cleveref}

\newcommand{\Pb}[1]{P_{\mathrm{#1}}}
\newcommand{\Tb}[1]{T_{\mathrm{#1}}}
\newcommand{\G}[1]{\Gamma_{\mathrm{#1}}}
\newcommand{\Lb}[1]{L_{\mathrm{#1}}}
\newcommand{\Ga}{\Gamma_{\mathrm{source}}}

\begin{document}
\title[REACH receiver design]{Receiver design for the REACH global 21-cm signal experiment}

\author*[1,2]{\fnm{Ian L. V.} \sur{Roque}}\email{ilvr2@cam.ac.uk}
\author[1]{\fnm{Nima} \sur{Razavi-Ghods}}
\author[1]{\fnm{Steven H.} \sur{Carey}}
\author[1]{\fnm{John A.} \sur{Ely}}
\author[1,2]{\fnm{Will} \sur{Handley}}
\author[3]{\fnm{Alessio} \sur{Magro}}
\author[4]{\fnm{Riccardo} \sur{Chiello}}
\author[1]{\fnm{Tian} \sur{Huang}}

\author[1]{\fnm{P.} \sur{Alexander}}
\author[1,2]{\fnm{D.} \sur{Anstey}}
\author[5,6,7]{\fnm{G.} \sur{Bernardi}}
\author[1,2]{\fnm{H. T. J.} \sur{Bevins}}
\author[8]{\fnm{J.} \sur{Cavillot}}
\author[9]{\fnm{W.} \sur{Croukamp}}
\author[1,2]{\fnm{J.} \sur{Cumner}}
\author[1,2]{\fnm{E.} \sur{de Lera Acedo}}
\author[9]{\fnm{D. I. L.} \sur{de Villiers}}
\author[2,10]{\fnm{A.} \sur{Fialkov}}
\author[1,2]{\fnm{T.} \sur{Gessey-Jones}}
\author[1]{\fnm{Q.} \sur{Gueuning}}
\author[1]{\fnm{A. T.} \sur{Josaitis}}
\author[11]{\fnm{G.} \sur{Kulkarni}}
\author[1,2]{\fnm{S. A. K.} \sur{Leeney}}
\author[1,2]{\fnm{R.} \sur{Maiolino}}
\author[12]{\fnm{P. D.} \sur{Meerburg}}
\author[11]{\fnm{S.} \sur{Mittal}}
\author[13,14]{\fnm{M.} \sur{Pagano}}
\author[9]{\fnm{S.} \sur{Pegwal}}
\author[9]{\fnm{C.} \sur{Pieterse}}
\author[15]{\fnm{J. R.} \sur{Pritchard}}
\author[12]{\fnm{A.} \sur{Saxena}}
\author[1,2]{\fnm{K. H.} \sur{Scheutwinkel}}
\author[1]{\fnm{P.} \sur{Scott}}
\author[1]{\fnm{E.} \sur{Shen}}
\author[13,14]{\fnm{P. H.} \sur{Sims}}
\author[6,7]{\fnm{O.} \sur{Smirnov}}
\author[16,17,18]{\fnm{M.} \sur{Spinelli}}
\author[3,4]{\fnm{K.} \sur{Zarb-Adami}}

\affil[1]{\orgdiv{Cavendish Astrophysics}, \orgname{University of Cambridge}, \orgaddress{\city{Cambridge}, \country{UK}}}
\affil[2]{\orgdiv{Kavli Institute for Cosmology}, \orgname{University of Cambridge}, \orgaddress{\city{Cambridge}, \country{UK}}}
\affil[3]{\orgdiv{Institute of Space Sciences and Astronomy}, \orgname{University of Malta}, \orgaddress{\city{Msida}, \country{Malta}}}

\affil[4]{\orgdiv{Physics Department}, \orgname{University of Oxford}, \orgaddress{\city{Oxford}, \country{UK}}}
\affil[5]{\orgdiv{Istituto di Radioastronomia}, \orgname{Istituto nazionale di astrofisica}, \orgaddress{\city{Bologna}, \country{Italy}}}
\affil[6]{\orgdiv{Department of Physics and Electronics}, \orgname{Rhodes University}, \orgaddress{\city{Grahamstown}, \country{South Africa}}}
\affil[7]{\orgname{South African Radio Astronomy Observatory}, \orgaddress{\city{Cape Town}, \country{South Africa}}}
\affil[8]{\orgdiv{Antenna Group}, \orgname{Universit\'e catholique de Louvain}, \orgaddress{\city{Louvain-la-Neuve}, \country{Belgium}}}
\affil[9]{\orgdiv{Department of Electrical and Electronic Engineering}, \orgname{Stellenbosch University}, \orgaddress{\city{Stellenbosch}, \country{South Africa}}}
\affil[10]{\orgdiv{Institute of Astronomy}, \orgname{University of Cambridge}, \orgaddress{\city{Cambridge}, \country{UK}}}
\affil[11]{\orgdiv{Department of Theoretical Physics}, \orgname{Tata Institute of Fundamental Research}, \orgaddress{\city{Mumbai}, \country{India}}}
\affil[12]{\orgdiv{Faculty of Science and Engineering}, \orgname{University of Groningen}, \orgaddress{\city{Groningen}, \country{Netherlands}}}
\affil[13]{\orgdiv{Trottier Space Institute}, \orgname{McGill University}, \orgaddress{\city{Montr\'eal}, \country{Canada}}}
\affil[14]{\orgdiv{Department of Physics}, \orgname{McGill University}, \orgaddress{\city{Montr\'eal}, \country{Canada}}}
\affil[15]{\orgdiv{Department of Physics}, \orgname{Imperial College London}, \orgaddress{\city{London}, \country{UK}}}
\affil[16]{\orgdiv{Osservatorio Astronomico di Trieste}, \orgname{Istituto nazionale di astrofisica}, \orgaddress{\city{Trieste}, \country{Italy}}}
\affil[17]{\orgdiv{Institute of Fundamental Physics of the Universe}, \orgaddress{\city{Trieste}, \country{Italy}}}
\affil[18]{\orgdiv{Department of Physics and Astronomy}, \orgname{University of the Western Cape}, \orgaddress{\city{Bellville}, \country{South Africa}}}

\abstract{We detail the REACH radiometric system designed to enable measurements of the 21-cm neutral hydrogen line. Included is the radiometer architecture and end-to-end system simulations as well as a discussion of the challenges intrinsic to highly-calibratable system development. Following this, we share laboratory results based on the calculation of noise wave parameters utilising an over-constrained least squares approach. For five hours of integration on a custom-made source with comparable impedance to that of the antenna used in the field, we demonstrate a calibration RMSE of 80 mK. This paper therefore documents the state of the calibrator and data analysis in December 2022 in Cambridge before shipping to South Africa.}


\maketitle
%\showthe\textwidth
% ==============================================================================
\section{Introduction}\label{intro}
The Radio Experiment for the Analysis of Cosmic Hydrogen (REACH) \citep{reach} is designed to measure the impact of the intergalactic medium (IGM) on the 21-cm neutral hydrogen line attributed to X-ray and UV emission from the first bright objects in the Universe \citep{furlanetto}. This “global” experiment focuses on detecting the spatial 21-cm cosmic signature which is orders of magnitude smaller than the bright foregrounds at frequencies in the region of 50--200MHz. As such, the experiment requires instrumental calibration of millikelvin-level accuracy to remove systematics that would ordinarily hinder such a measurement.  

A number of global experiments have already been conducted in this domain such as SARAS \citep{saras} and LEDA \citep{leda} as well as EDGES, which in 2018 reported the detection of an absorption profile at 78 MHz, potentially revealing the general characteristics of the Epoch of Reionisation (EoR) and Cosmic Dawn such as the onset of reionisation and the start of active black hole accretion \citep{edgesNature}. While centred within the low-frequency radio regime proposed by theorists \citep{21cmTheory}, the signal depth is more than two times larger than predictions ($>0.5$ K), which if physical could suggest additional cooling of interstellar gas \citep{edgesNewPhysics} or an excess radio background \citep{excessRadioBackground}. Following the EDGES result, several studies have now disputed the findings such as SARAS-3 which rejects the EDGES best-fit profile with 95.3\% confidence \citep{sarasDisputeEdges}, hypothesising that analysis is still systematics dominated \citep{2018Natur.564E..32H}.

In response to the questioned reliability of the EDGES detection, REACH is designed to address some of the perceived limitations of previous experiments through re-evaluation of the ethos taken for data analysis and systematic modelling as well as placing emphasis on radio-frequency (RF) system stability and temperature control. Our approach includes high-quality RF components such as a well-matched low noise amplifier (LNA), switching electronics and calibration loads to enable measurements of noise wave parameters as linear analogues to the standardised noise parameters described in \citet{meys} which specify the noise generated and reflected by the first amplifier down to millikelvin levels. These considerations result in a fully automated in-field calibration system for determining the instrument characteristics with minimal human interaction.

The sensitive nature of the measurements performed by this radiometer necessitates that everything from the antenna terminals to the back-end digital system be well calibrated. In this paper, we detail the general calibration formalism (\cref{calibration}), followed by the full radiometric system design in \cref{radiometer}. Important corrections to be applied to the data prior to computing the calibration parameters are given in \cref{methods}. This is followed by results obtained from a least squares solver for a laboratory dataset taken over five hours in \cref{results} where we achieve an RMSE of approximately 80 mK for the “simulated” antenna and approximately 30 mK for long cable sources. We conclude with some lessons learned from the system design and results in \cref{conclusions}.  

This paper documents the state of the calibrator and associated data analysis before leaving the Cavendish Laboratory in Cambridge in December 2022. Further work will detail any adjustments in the instrument arising over 2023 from the travel to Stellenbosch in South Africa, through EMI testing and then onto the REACH site in the Karoo desert.

% ==============================================================================
\section{Calibration Formalism}\label{calibration}
\begin{figure}
    \includegraphics[width=\columnwidth]{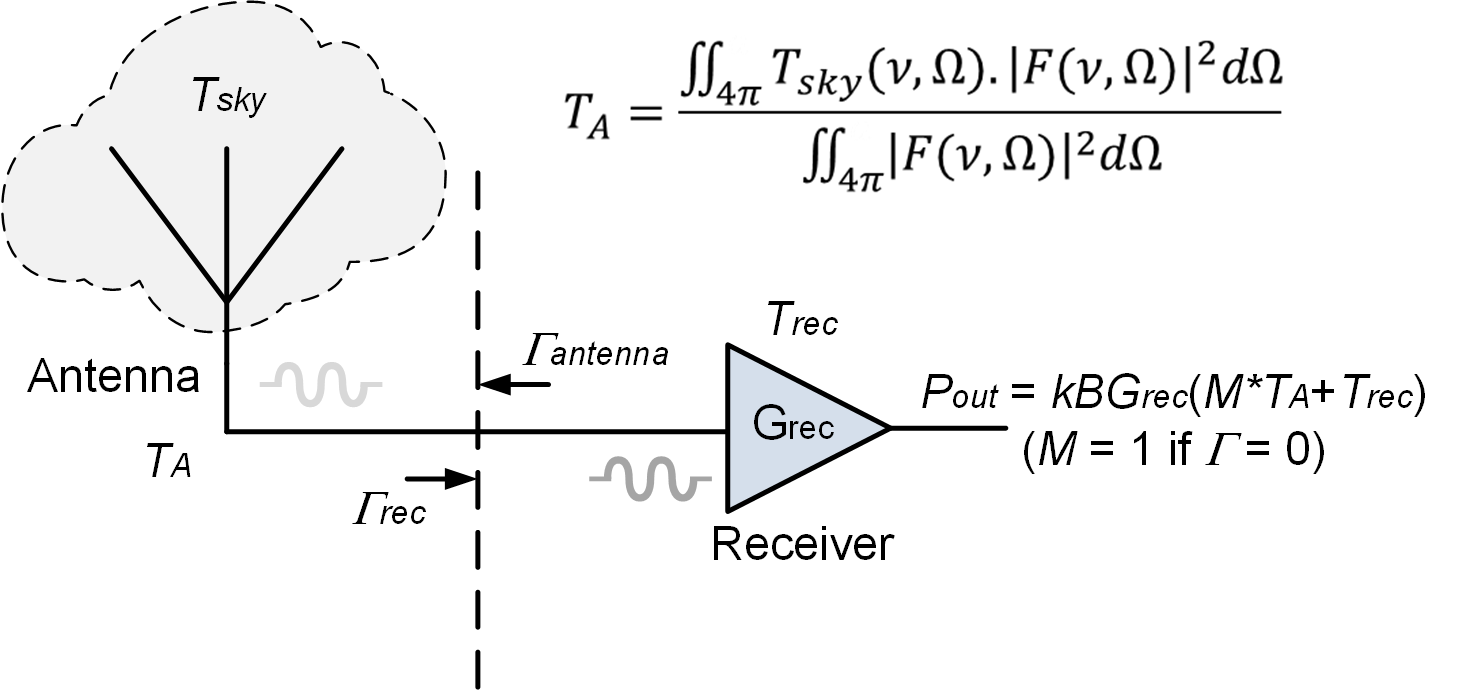}
    \caption{Simplified antenna-receiver interaction for a global experiment showing $\Tb{A}$ as the antenna beam $\lvert F \rvert^2$ integrated with the sky brightness function $\Tb{sky}$ over frequency $\nu$ and coordinates above the horizon $\Omega$ \citep{ant_temp}. Here, the receiver input reference plane is shown by the dashed line where antenna mismatches ($M \neq 1$) cause reflections that can enter the receiver as noise. Properties of the receiver such as its characteristic gain ($G_{\mathrm{rec}}$) contribute directly to the power spectral output, $P_{\mathrm{out}}$ measured by the back-end. This diagram assumes no antenna losses.}
    \label{fig:radiometer}
\end{figure}

The primary goal of the REACH radiometer is to effectively model the measurement system and remove systematics downstream of the antenna that hinder our ability to detect the cosmological signal. Whilst conceptually the REACH instrument relies only on a few subsystems, namely the antenna, receiver (front-end) and readout system (back-end), understanding the interaction between these subsystems (particularly the antenna and receiver) is critical in determining the absolute systematic structure to be removed from the data prior to any data analysis. A simplified picture of the radiometer problem is illustrated in \cref{fig:radiometer}, which shows the antenna and receiver (in this case denoting the entire RF signal chain).  

Of the effects shown in \cref{fig:radiometer}, the spectral shape of noise arising from mismatches ($M$) between the antenna and receiver is of primary importance for this work. If both subsystems are perfectly matched, the reflection coefficient is zero and the reliance on this form of calibration diminishes. However, it is simply impractical to develop an achromatic antenna that provides a perfect match to the receiver across a broad bandwidth as used by REACH to leverage spectral differences between astrophysical foregrounds and potential cosmological signals \citep{reach}. This limitation is often due to the practicalities of designing low-frequency antennas. Given that the foregrounds at these frequencies ($< 200$ MHz) are up to five orders of magnitude larger than the theoretical cosmic signature, any subsystem mismatch can result in considerable spectral variation across the measurement band on the order of tens of Kelvin. Furthermore, whilst historically relative calibration was used for narrow-band radiometers, wide-band instruments must obtain an absolute flux scale across the frequency band in order to measure the frequency-dependent sky-averaged brightness temperature \citep{rogersCal}.

A first-order relative spectral calibration of such a system is achieved through the use of three-position Dicke switching \citep{dicke} where, in addition to making antenna power spectral density (PSD) measurements $\left(\Pb{source}\right)$, PSDs are also obtained from a high-quality noise source $\left(\Pb{NS}\right)$ and an ambient 50 $\Omega$ load $\left(\Pb{L}\right)$ at a fixed input reference plane to the receiver. A preliminary antenna temperature $\Tb{source}^*$ can then be calculated using
\begin{equation}
    \label{eqn:tcalstar}
    \Tb{source}^* = \Tb{NS} \left(\frac{\Pb{source}-\Pb{L}}{\Pb{NS}-\Pb{L}}\right) + \Tb{L},
\end{equation}
with the noise wave parameters $\Tb{L}$ and $\Tb{NS}$ relating to the noise temperatures of the load and the excess noise temperature of the noise source, respectively. Calculation of the preliminary antenna temperature serves to calibrate out time-dependent system gain ($g_{\mathrm{sys}}$) arising from the components within the receiver chain \citep{edgesCal}. 

To model the system interactions, we first define the PSDs obtained for the reference load and noise source. Since the reflection coefficients of these components are generally measured to be very small (typically on the order of 0.005 or less), we can simplify these interactions by assuming the reflection coefficient is zero, resulting in 
\begin{equation}
    \label{eqn:pl}
    \Pb{L} = g_{\mathrm{sys}} \left[\Tb{L}\left(1-\lvert \G{rec}\rvert ^2\right)+\Tb{0}\right],
\end{equation}

\begin{equation}
    \label{eqn:pns}
    \Pb{NS} = g_{\mathrm{sys}} \left[\left(\Tb{L}+\Tb{NS}\right)\left(1-\lvert\G{rec}\rvert^2\right)+\Tb{0}\right].
\end{equation}
Here $\G{rec}$ is the reflection coefficient of the receiver. $g_{\mathrm{sys}}$ and $\Tb{0}$ are the system gain and temperature offset, respectively \citep{edgesCal}. The Boltzmann constant as shown in \cref{fig:radiometer} has been truncated into $g_{\mathrm{sys}}$ for the power spectral density equations as the terms cancel when dividing the PSDs.

In the case of the source (calibrator or antenna), the assumption of a low reflection coefficient is no longer true. Therefore we can use the following definition \citep{roque, edgesCal}
\begin{equation}
    \label{eqn:pant}
    \begin{aligned}
        \Pb{source} = g_{\mathrm{sys}} \Bigg[ &\Tb{source}\left(1-\lvert\Ga\rvert^2\right)\left\lvert\frac{\sqrt{1 - \rvert\G{rec}\lvert^2}}{1-\Ga\G{rec}}\right\rvert^2 \\
        + & \Tb{unc}\lvert\Ga\rvert^2\left\lvert\frac{\sqrt{1 - \rvert\G{rec}\lvert^2}}{1-\Ga\G{rec}}\right\rvert^2 \\
        + & \Tb{cos}\operatorname{Re}\left(\Ga\frac{\sqrt{1 - \lvert\G{rec}\rvert^2}}{1-\Ga\G{rec}}\right) \\
        + & \Tb{sin}\operatorname{Im}\left(\Ga\frac{\sqrt{1 - \lvert\G{rec}\rvert^2}}{1-\Ga\G{rec}}\right) 
        + \Tb{0} \Bigg].
    \end{aligned}
\end{equation}

Here, $\Tb{source}$ is our calibrated input temperature and $g_{\mathrm{sys}}$ is the system gain referenced to the receiver input. Since the reference plane in our system is fixed, we can assume both $g_{\mathrm{sys}}$ and $\Tb{0}$ are the same as in equations \cref{eqn:pl} and \cref{eqn:pns} which simplifies our calibration equation later. $\Tb{unc}$, $\Tb{cos}$, and $\Tb{sin}$ are the noise wave parameters introduced by \citet{meys} and \citet{rogersCal} to calibrate the instrument. $\Tb{unc}$ represents the portion of noise reflected by the antenna that is uncorrelated with the output noise of the LNA, whilst $\Tb{cos}$ and $\Tb{sin}$ are the portions of reflected noise correlated with noise from the LNA \citep{rogersCal, roque}. In the EDGES experiment, these calibration quantities are modelled using seven-term polynomials in frequency \citep{edgesCal}.

Inserting the definitions for $\Pb{source}$, $\Pb{L}$ and $\Pb{NS}$ into \cref{eqn:tcalstar} yields our calibration equation which relates the noise wave parameters to measured quantities of our system.  All parameters are frequency-dependent.
\begin{equation}
    \label{eqn:caleqn}
    \begin{aligned}
        \Tb{NS}\left( \frac{\Pb{source} - \Pb{L}}{\Pb{NS} - \Pb{L}} \right) + \Tb{L}&= \Tb{source}\left[ \frac{1-\lvert\G{source}\rvert^2}{\lvert1-\Ga\G{rec}\rvert^2} \right] \\
        & + \Tb{unc}\left[ \frac{\lvert\Ga\rvert^2}{\lvert1-\Ga\G{rec}\rvert^2} \right] \\
        & + \Tb{cos}\left[ \frac{\operatorname{Re}\left(\frac{\Ga}{1-\Ga\G{rec}}\right)}{\sqrt{1-\lvert\G{rec}\rvert^2}} \right] \\
        & + \Tb{sin}\left[ \frac{\operatorname{Im}\left(\frac{\Ga}{1-\Ga\G{rec}}\right)}{\sqrt{1-\lvert\G{rec}\rvert^2}} \right]. \\ 
    \end{aligned}
\end{equation}

We can then rewrite \cref{eqn:caleqn}, separating out the measured quantities (X-terms) as detailed in \citet{roque}, resulting in a simplified form
\begin{equation}
    \Tb{source} = X_{\mathrm{unc}}\Tb{unc} + X_{\mathrm{cos}}\Tb{cos} + X_{\mathrm{sin}}\Tb{sin} + X_{\mathrm{NS}}\Tb{NS} + X_{\mathrm{L}}\Tb{L}.
\end{equation}
Furthermore, the linear form of this equation allows us to segregate the instrument measurements and models of our noise wave parameters into separate matrices
\begin{align}\label{eqn:theta}
    \mathbf{X} &\equiv \begin{pmatrix} 
    X_\mathrm{unc} \quad 
    X_\mathrm{cos} \quad
    X_\mathrm{sin} \quad
    X_\mathrm{NS} \quad
    X_\mathrm{L} \end{pmatrix},\nonumber\\
    \boldsymbol{\Theta} &\equiv \begin{pmatrix} 
    T_\mathrm{unc}\quad
    T_\mathrm{cos}\quad
    T_\mathrm{sin}\quad
    T_\mathrm{NS}\quad
    T_\mathrm{L}\end{pmatrix}^\top,
\end{align}
which condenses our calibration equation, with noise term $\sigma$, to
\begin{equation}\label{eqn:linearmodel}
    \mathbf{T}_\mathrm{source} = \mathbf{X}\boldsymbol{\boldsymbol{\Theta}}+\sigma.
\end{equation}

This equation can be solved to determine $\boldsymbol{\Theta}$, representing the five noise wave parameters, in a number of ways. As described in \citet{roque}, a Bayesian framework has already been developed and tested on simulated data. However, we have also developed a standard least squares method that computes the calibration coefficients on a frequency-by-frequency basis rather than one using polynomial fitting. It is the latter method that we will rely on to show the functionality of this system in \cref{results}, however, the different approaches being proposed for analysing this data will be addressed in a later paper.

The system is calibrated by solving the set of linear \cref{eqn:caleqn}’s for our noise wave parameters with data from various simple ‘calibrator’ devices informing the solution. The number of calibrators can vary as long as there are a sufficient number of devices to constrain the equations under the caveat that these sources have distinct impedances to give maximal information on the response of the system. Once a solution is calculated, this information is applied to a complex impedance of unknown temperature such as an antenna configuration looking at the night sky. The procedure for choosing calibrators under the REACH experiment can be found in \cref{sec:sources}.

% ==============================================================================
\section{Receiver design}\label{radiometer}
One approach setting the REACH radiometer apart from other systems targeting 21-cm cosmology is the capability of in-field calibration using minimal laboratory-based data. The reason for this is that once the system is deployed and exposed to the elements, ensuring environmental stability over long time periods is difficult. This necessitates an effective, fully autonomous system for data acquisition on a regular basis to be used for routine updates or calculation of the noise wave parameters.  

\begin{figure}
    \includegraphics[width=\columnwidth]{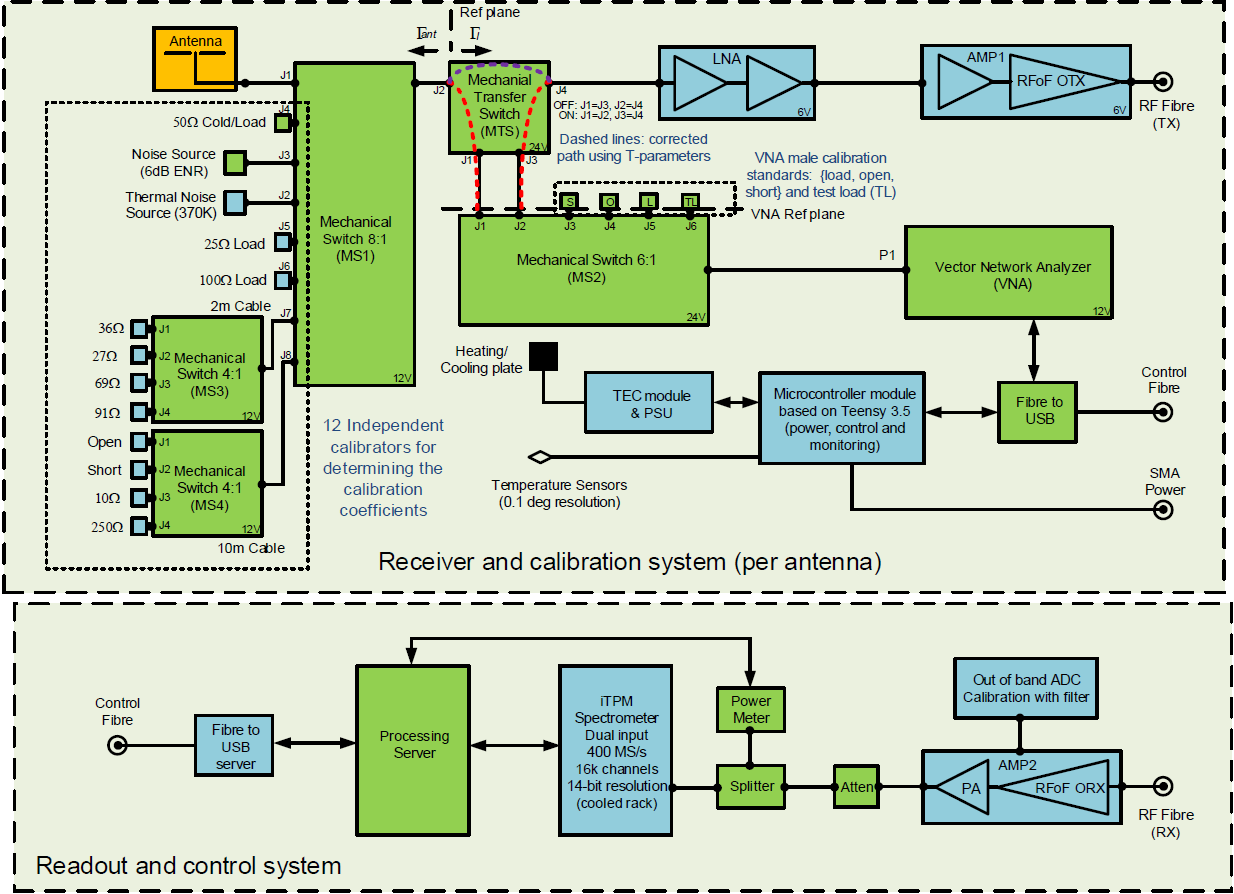}
    \caption{An overview of the REACH radiometer showing calibration sources and the antenna connected to an 8-way mechanical input switch which is then connected to the receiver. The green sub-blocks represent off-the-shelf components, whilst blue represent custom designs. $\Gamma_{ant}$ represents the reflection coefficient of the antenna or calibrator, $\Gamma_{l}$ is the reflection coefficient of the receiver. The red dashed line represents the extra path measured by the VNA that is not present during spectral measurements while the purple dashed line is the path present exclusively during spectral measurements. Corrections for these additional paths are detailed in \cref{sparams}. ‘ENR’ is the Excess Noise Ratio of a Noisecom NC346A noise source; ‘OTX’ indicates an optical transmitter; ‘TX’ indicates transmission mode; ‘TEC’ stands for Thermoelectric Cooling; ‘SMA’ is a SubMiniature version-A connector; ‘PA’ is Power Amplifier; ‘RX’ indicates reception mode and ‘Atten.’ represents a signal attenuator. Updated from figure included in \citet{reach}.}
    \label{fig:overview}
\end{figure}

The REACH system relies on three forms of data to calibrate, which are measured by different circuits. The reflection coefficients of the calibration sources, antenna and receiver are measured by a Copper Mountain Technologies TR1300/1 vector network analyser (VNA). Power spectral densities are measured by a SanitasEG \emph{italian} Tile Processor Module (iTPM) \citep{itpm} spectrometer and finally, the physical temperature of the sources are measured by a Pico Technology TC-08 data-logging thermocouple. An overview of the radiometer is shown in \cref{fig:overview}.

The radiometer front-end houses the main receiver and calibration sub-system while the back-end, separated by a 100 metre distance, handles data collection, control and signal processing. As shown in \cref{fig:overview}, the front-end employs low-loss mechanical switches (typically 0.01 dB in this band) with better than 100 dB isolation. The main 8-way switch (MS1) allows switching between the antenna and various sources described in \cref{sec:sources}. A transfer switch (MTS) permits VNA measurement of the source and LNA reflection coefficients. To initially calibrate the VNA, switch MS2 toggles between a short (S), open (O) and load (L) standard before verifying the calibration accuracy against an independently characterised test load (TL). Since this VNA calibration is done at a different reference plane, calculations are performed to de-embed the extra signal path and ‘move’ the VNA data to the receiver reference plane as described in \cref{methods}.

An onboard microcontroller unit facilitates switching along with other functions such as environmental temperature control. A USB-to-fibre converter is used to send signals to the microcontroller and VNA. Following the LNA, another module (AMP1) amplifies and filters the signal before transmission via a radio-frequency-over-fibre (RFoF) optical transmitter. The RF optical signal is transmitted via single-mode optical cables to avoid interference and to limit signal loss. All signalling, whether control or RF, is transmitted via single-mode fibre back to the back-end node where they are converted to electrical signals. In the RF signal chain, AMP2 is used to convert back to RF, offering further filtering and amplification prior to digitisation in the readout system (iTPM) controlled by a server. Further details of the full environmental node and other control aspects are discussed in \citet{reach} but are outside the scope of this paper. Essential design blocks are detailed in the following sections.

\subsection{Calibration sources}\label{sec:sources}
One of the critical elements of the REACH radiometer is the calibration sources. The primary objective of using these sources is to permit strategic sampling of the noise waves as a function of impedance. In the case of EDGES, four sources were used; a heated (hot) and ambient (cold) 50 $\Omega$ load were measured to obtain a scale and offset, denoted as $C_1$ and $C_2$ in EDGES terminology, respectively. This was followed by two additional calibrators made from coaxial cables terminated with a shorted load and an open load which provide information on the noise wave parameters used to calibrate the instrument \citep{edgesCal}.

For REACH, however, we can rely on up to 12 calibrators as shown in \cref{fig:overview}. These are in addition to the reference sources which are used to obtain $\Pb{NS}$ and $\Pb{L}$. The final calibrators used are listed below.
\begin{itemize}
    \item A thermal noise source (50 $\Omega$ heated to 370 K)
    \item An ambient 50 $\Omega$ load (the same load used to obtain $\Pb{L}$)
    \item Ambient 25 $\Omega$ and 100 $\Omega$ loads
    \item A 2 m cable connected to switch MS3 (terminated in 27 $\Omega$, 36 $\Omega$, 69 $\Omega$, or 91 $\Omega$) at ambient temperature
    \item A 10 m cable connected to switch MS4 (terminated in Open, Short, 10 $\Omega$, or 250 $\Omega$) at ambient temperature
\end{itemize}

\begin{figure}
    \includegraphics[width=\columnwidth]{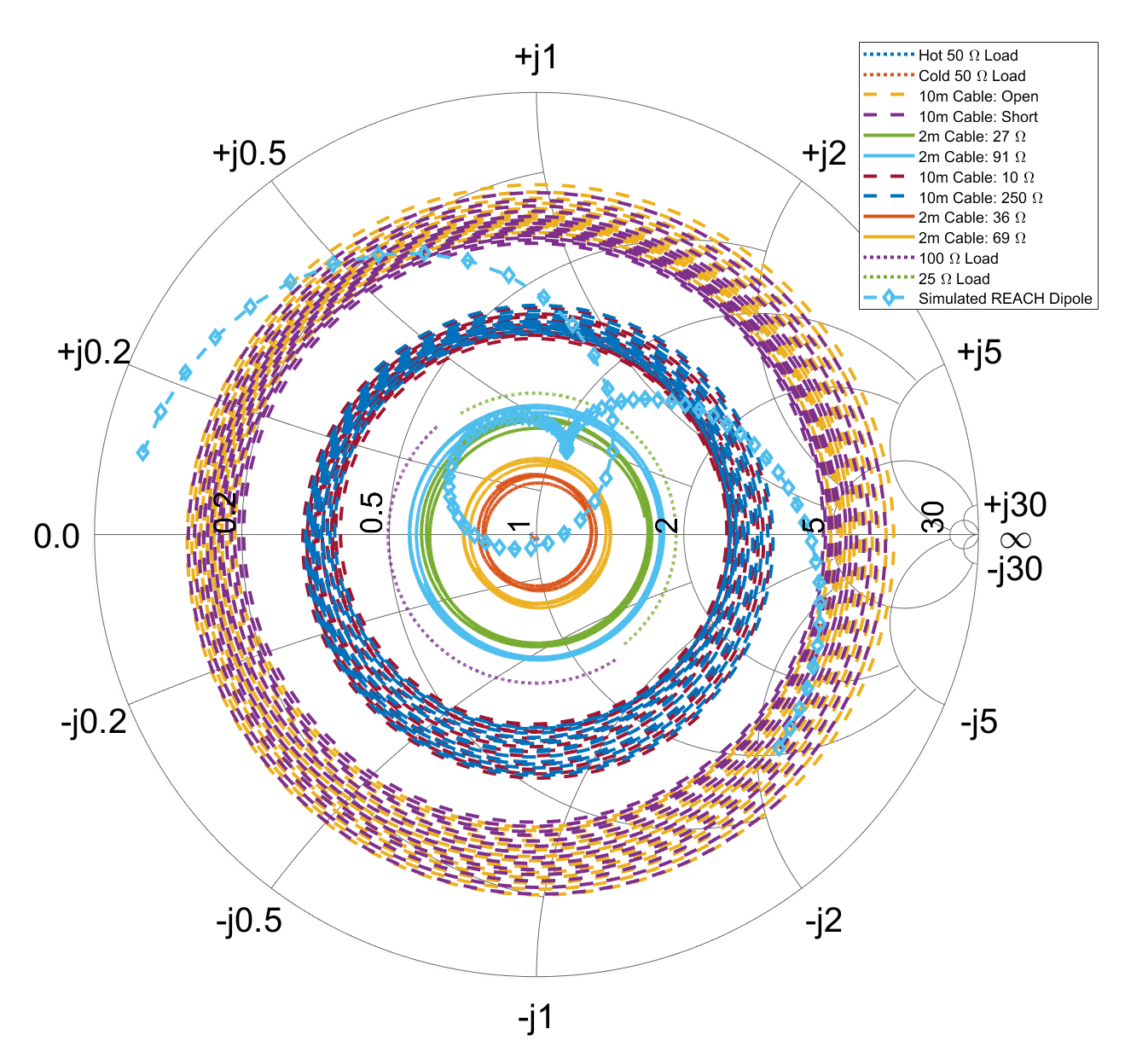}
    \caption{Smith Chart showing the impedance of twelve calibration sources and the simulated REACH antenna (internal variant \#0744) with the centre of the plot indicating an impedance of 50 $\Omega$. The plot ranges from 50--150 MHz with the antenna curve starting at 50 MHz on the left-hand side. The extensive coverage in impedance space by our calibration standards can be interpreted as a substantial amount of information regarding the characteristic response of the instrument. Note that the impedances of the ambient and heated 50 $\Omega$ loads lie directly in the middle of the chart and are partially obscured by the antenna plot. Updated from figure included in \citet{reach}.}
    \label{fig:smith}
\end{figure}

A diverse set of calibration sources will give the maximal amount of information for calibrating the receiver. \Cref{fig:smith} demonstrates the comprehensive scope of frequency-dependant impedances for our calibration sources as well as a simulated impedance of the REACH dipole antenna covering 50--150MHz \citep{cumner}. As noted in \citet{reach}, REACH will also use a log periodic spiral antenna to make concurrent observations from 50--170MHz. Furthermore, since we are trying to determine five frequency-dependent noise wave parameters (denoted as $\boldsymbol{\Theta}$), it helps to have access to more than four calibration sources, over-constraining the parameters in a frequency-by-frequency least squares sense. \Cref{fig:smith} also demonstrates measurements of the 25 $\Omega$ and 100 $\Omega$ loads as half circles on the Smith chart, which differs from the theoretical points at 25 $\Omega$ and 100 $\Omega$ due to the practical limitations of real-world impedance measurement and exacerbated by the additional RF path in our receiver between the MS1 switch and the VNA reference plane as shown in \cref{fig:overview}. These effects were the motivation for the corrections detailed in \cref{sparams}.

Of the 12 calibrators accessed by the radiometer, the heated load determines an absolute temperature scale. It is constructed from a 50 $\Omega$ load, heated with a proportional heater to 370 K and connected to MS1 via a thin 4-inch coaxial cable. The construction of the heated load module, as a 50 $\Omega$ resistor connected to a heating element directly monitored by a thermocouple, yields accurate measurement of the heated load temperature\footnote{Sum of $\pm0.2$\% of reading and $\pm0.5$ K based on Pico Technology TC-08 specifications}. This is beneficial for the removal of systematic noise via accurate noise wave parameter derivation, but sacrifices the constant noise power in frequency native to the diode noise source. This constant noise power is necessary for maximal radiometer measurement accuracy through removal of the time-dependent system gain fluctuations via the Dicke switching procedure.

\subsection{RF signal chain}
\begin{figure}
    \includegraphics[width=\columnwidth]{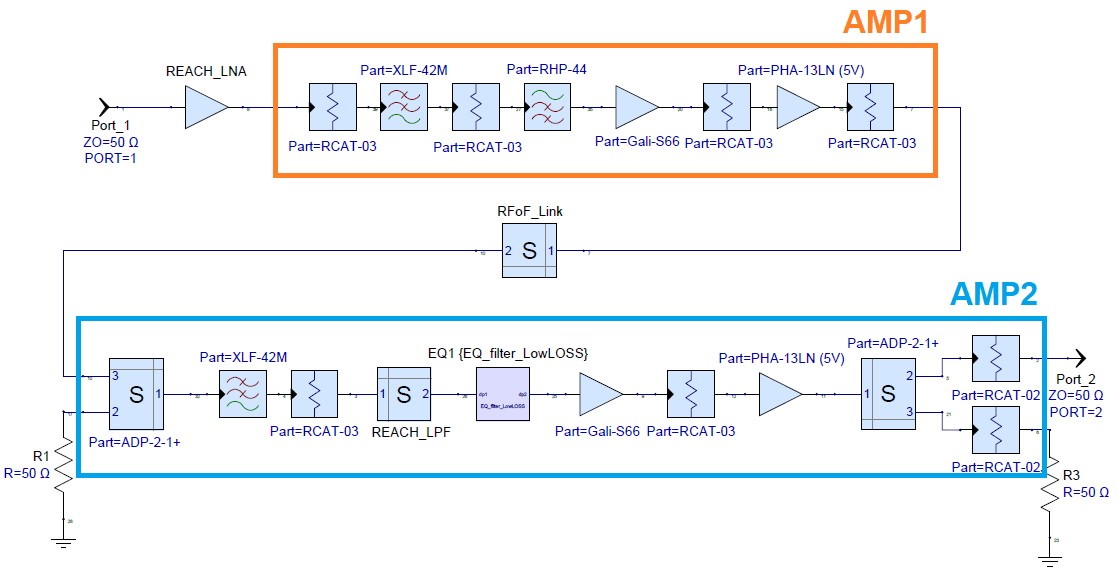}
    \caption{REACH RF system end-to-end block simulations made with the Keysight PathWave RF Synthesis software showing the LNA, as well as custom design for AMP1 and AMP2.}
    \label{fig:RF_system}
\end{figure}

The RF signal chain shown in \cref{fig:overview} consists of three custom-designed components: LNA, AMP1 and AMP2. These components have been simulated in Keysight PathWave RF Synthesis (Genesys) software, relying especially on linear analysis as well as the Spectrasys RF Systems software for RF budget simulations. The optimisation tool has also been used for tuning, in particular for filter design. Many of the amplifier components used in the simulation have either been measured directly with a VNA or modelled using substrate-scalable components developed by Modelithics. An overview of the RF simulation setup is shown in \cref{fig:RF_system}.

\subsubsection{LNA}
The LNA is designed using a pair of cascaded CMA-84+ SMT gain blocks from Mini-Circuits and high-quality attenuator blocks to achieve exceptional input matching and a spectrally flat passband response. The LNA is not especially low noise in the strict sense as it has a flat noise figure of 5.1 dB. However it is expected that this will have a limited impact on the REACH global experiment which is not sensitivity limited. In the range where the REACH dipole is best matched (60--120 MHz), the system will be sky noise dominated, whilst at frequencies greater than 120 MHz, we expect reduced sensitivity. The typical trade-off made with such amplifiers is noise versus match. Typically the better the match response of the amplifier, the poorer the noise figure, although these can be tuned to a certain degree if relying on discrete components.

\begin{figure}
    \includegraphics[width=\columnwidth]{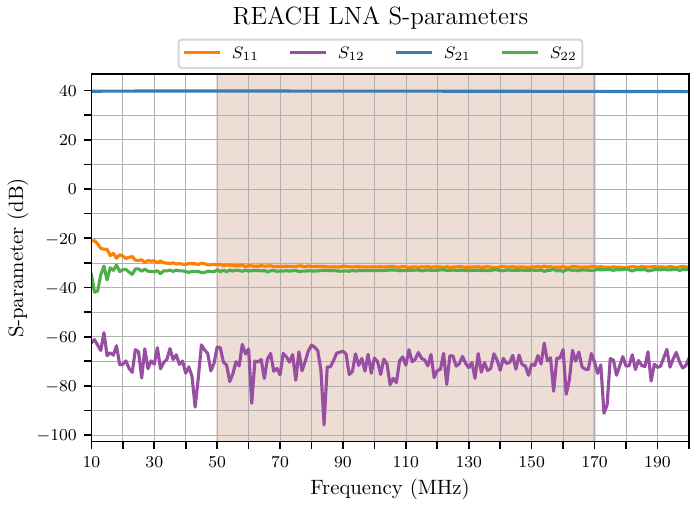}
    \caption{Measured S-parameters of the REACH LNA imported to the end-to-end simulation diagrammed in \cref{fig:RF_system}. The shaded region represents the REACH observation band of 50--170 MHz. The $S_{11}$ and $S_{22}$ are shown to have a good match at -30 dB across the observation band while the $S_{21}$ demonstrates remarkable stability. Adapted from figure included in \citet{reach}.}
    \label{fig:sim_a}
\end{figure}

For REACH, the main priority is reducing the amplifier input reflection coefficient ($S_{11}$) to -30 dB or lower since this would reduce the impact of the noise waves. Furthermore, another important consideration was gain variation with temperature. In both cases, having evaluated a number of different amplifiers, we settled on the CMA-84+. Having higher noise from the LNA directly impacts the noise waves we obtain resulting in larger values for $\Tb{unc}$, $\Tb{cos}$, and $\Tb{sin}$ which is further amplified by the antenna reflection coefficient. In general, having smaller noise waves is more optimal, however, having tested the calibration system against two different LNA modules including an amplifier based on an ERA-50SM+ chip offering a NF of 3.3 dB, we determined better stability over time with the CMA-84+. Measurements of the LNA S-parameters are included in \cref{fig:sim_a} showing a good input-output match at better than -30 dB over the observational band as well as a remarkably flat response of 40 dB of gain.

\subsubsection{AMP1 and AMP2}
Following the LNA, the RF signal is further amplified and high-pass filtered in the AMP1 module before going through a passive 1310 nm RFoF link. Since the link has a loss of typically 18 dB, mainly constrained by the Relative Intensity Noise (RIN) of the laser in the optical transmitter, a reasonable amount of upfront gain (approximately 70 dB) is used to reduce the impact of higher noise on the system. In the simulations shown in \cref{fig:RF_system}, the RFoF link consists of the optical transmitter (in AMP1), a length of 100 m single-mode fibre and the optical receiver (in AMP2). This was characterised by a VNA at different power levels and used as a single block in the full end-to-end simulations.

The RFoF link minimises the impact of radio-frequency interference (RFI) and cable loss over the 100 m separation between the front- and back-ends compared to coaxial cables. At 1310 nm, the RF loss in the single-mode cable is typically less than 1 dB including the connections at either end. The RFoF module itself was designed by Polycom using our specifications for the HERA EoR experiment \citep{hera}. The optical transmitter and receiver sub-assemblies were small solderable printed circuit boards terminated in FC/APC connections at the end of a 0.5 m pigtail.

In the back-end, the AMP2 module was used to convert the optical signal back to RF, providing further filtering and amplification. A custom-designed 11-order Cauer Chebyshev low-pass filter was used to sharply filter signals above 170 MHz, since the goal was to use an RF sampling of 400 MSPS in the iTPM, limiting out-of-band signal power. Furthermore, a 2-stage MMIC reflectionless low pass filter from Mini-Circuits (XLF-42M+) was used in both AMP1 and AMP2 to filter much higher frequency out-of-band signals up to many GHz. To flatten the passband to 2 dB, an additional low-loss 3 dB equalisation circuit was used in AMP2. Both the AMP1 and AMP2 units rely on the GALI-S66+ limiting amplifier and the PHA-13LN+ mid-power amplifier to achieve the best dynamic range prior to the analog-to-digital converter (ADC) in the iTPM.

\begin{figure}
    \includegraphics[width=\columnwidth]{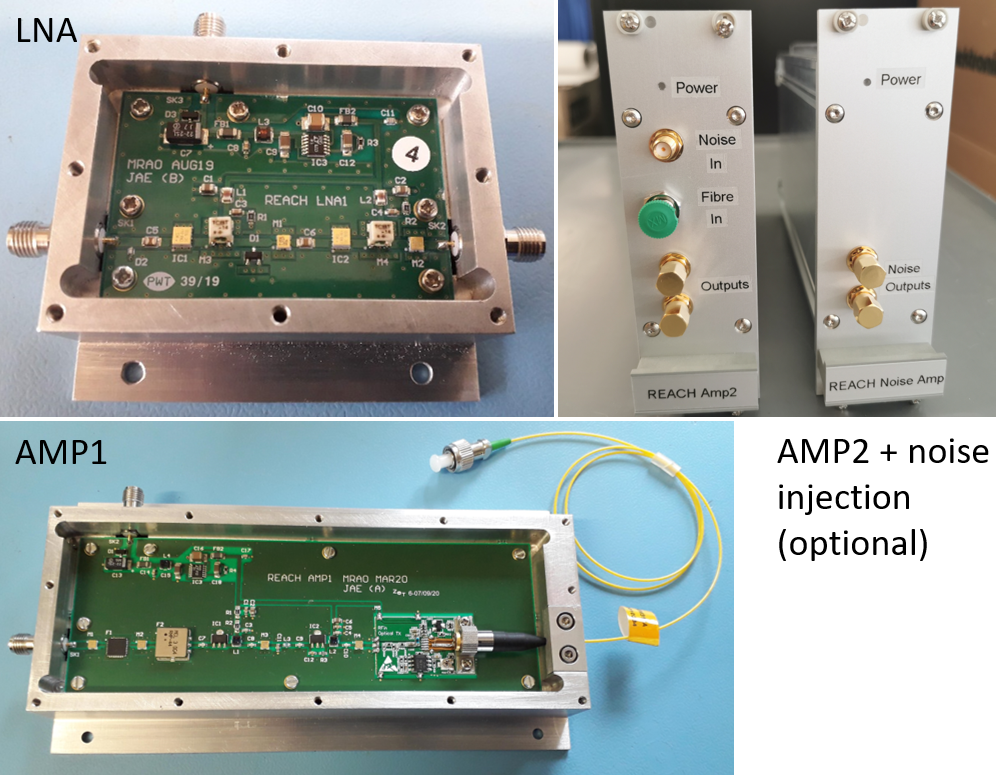}
    \caption{REACH RF chain hardware including the LNA and AMP1 interiors as well as the back-end AMP2. Displayed on the far right is the out-of-band noise injection module for conditioning of the ADC.}
    \label{fig:RF_chain}
\end{figure}

 AMP2 also permits the use of an out-of-band signal injection (continuous wave or filtered noise) to condition the ADC although this was not used in the final system since it offered minimal improvements to the data. AMP2 has the capability of outputting two equal signals via a well-balanced power splitter, with the second output either going to another ADC path or a separate power meter for signal monitoring. \Cref{fig:RF_chain} shows all the components used in the RF path including the optional out-of-band noise injection module which is band-limited to DC--20 MHz.

\begin{figure}
    \includegraphics[width=\columnwidth]{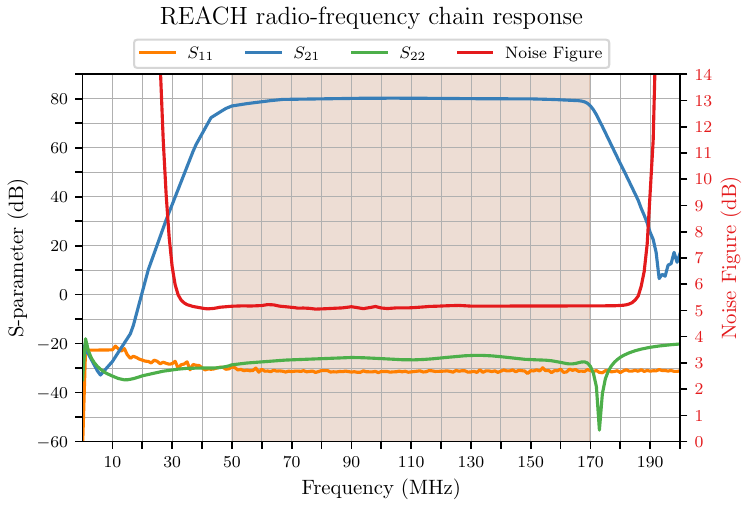}
    \caption{Simulated radio-frequency response of the REACH end-to-end signal-chain as diagrammed in \cref{fig:RF_system} which includes the LNA, AMP1 and AMP2. The shaded region represents the REACH observation band where we see a flat noise figure throughout. Adapted from figure included in \citet{reach}.}
    \label{fig:sim_b}
\end{figure}

A simulation of the full RF chain end-to-end system response from Keysight's PathWave RF Synthesis (Genesys) software using Modelithics substrate scalable models and the measured LNA data is shown in \cref{fig:sim_b}. In this analysis, each block in the RF chain was first simulated, then built and measured with a VNA with the filtered passband response optimised through the tools available in Genesys.

\subsection{Microcontroller unit}
To achieve autonomous operation in the field given the space constraints of the front-end enclosure and the requirement for low noise, a decision was made early on in the project to develop a custom microcontroller unit which would form the heart of the radiometer. The REACH microcontroller unit is based on a Teensy 3.5, which is a tiny feature-packed development board designed by PJRC and pre-flashed with a bootloader. This microcontroller allows easy programming of various functions needed in the front-end including control of switches and additional temperature monitoring. A set of low-level functions were developed and used in this project.\footnote{More details can be found at \url{https://gitlab.developers.cam.ac.uk/phy/ra/shc44/reach-rx/-/tree/main/controller}}

\begin{figure}
    \includegraphics[width=\columnwidth]{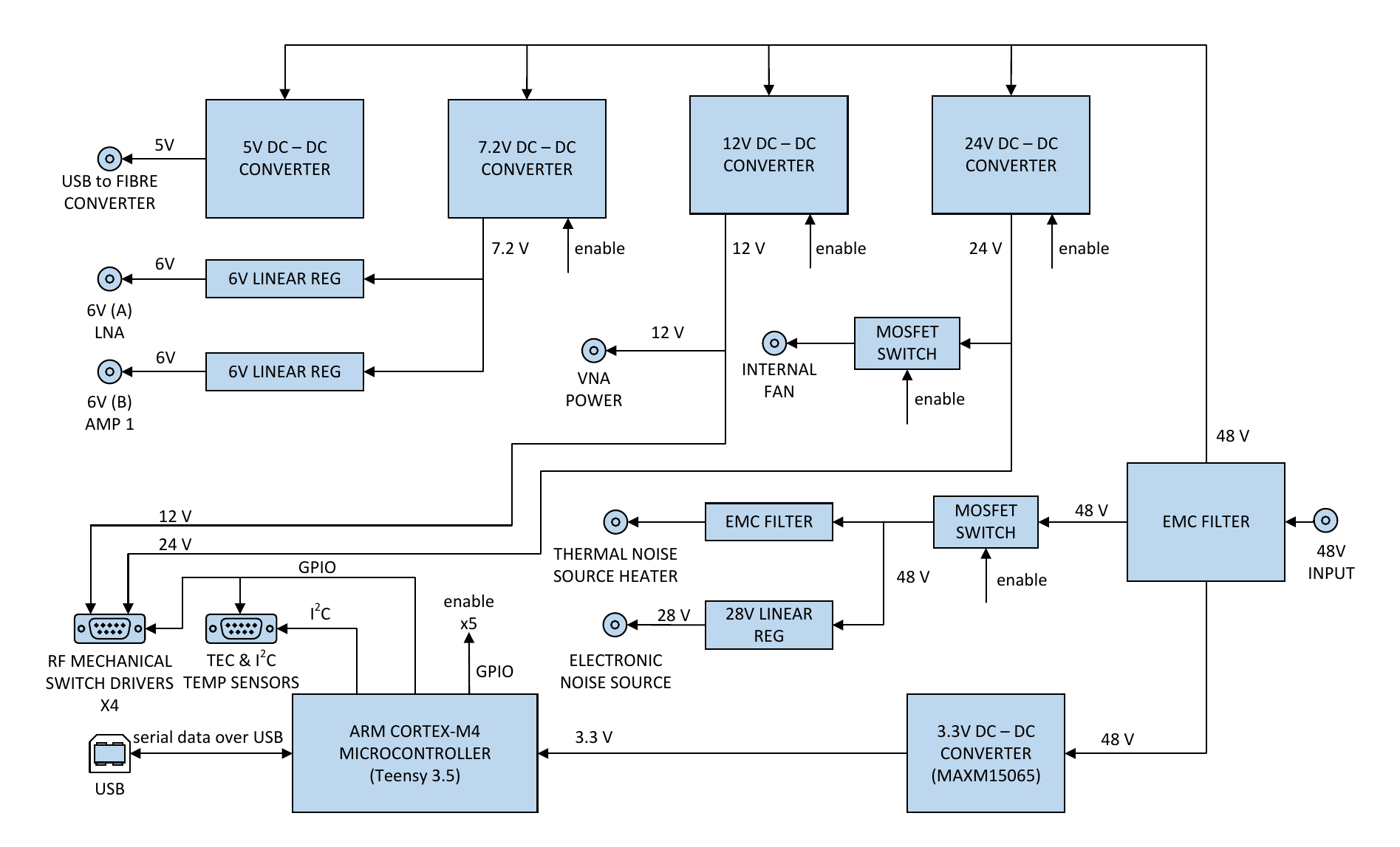}
    \caption{Detailed microcontroller block diagram showing the components, connections and power considerations incorporated into the design.}
    \label{fig:uc_detail}
\end{figure}

The microcontroller unit provides power supplies for everything except thermal management. A high level of DC filtering is provided on the input supply to the receiver (typically 48 V) and for noise critical supplies (LNA, AMP1), a combination of SMPS and linear regulators are employed for an optimum combination of efficiency and low noise. With all supplies on and fully loaded, the temperature rise inside the microcontroller enclosure is only 2 K. A detailed block diagram of the microcontroller unit is shown in \cref{fig:uc_detail}.

\begin{figure}
    \includegraphics[width=\columnwidth]{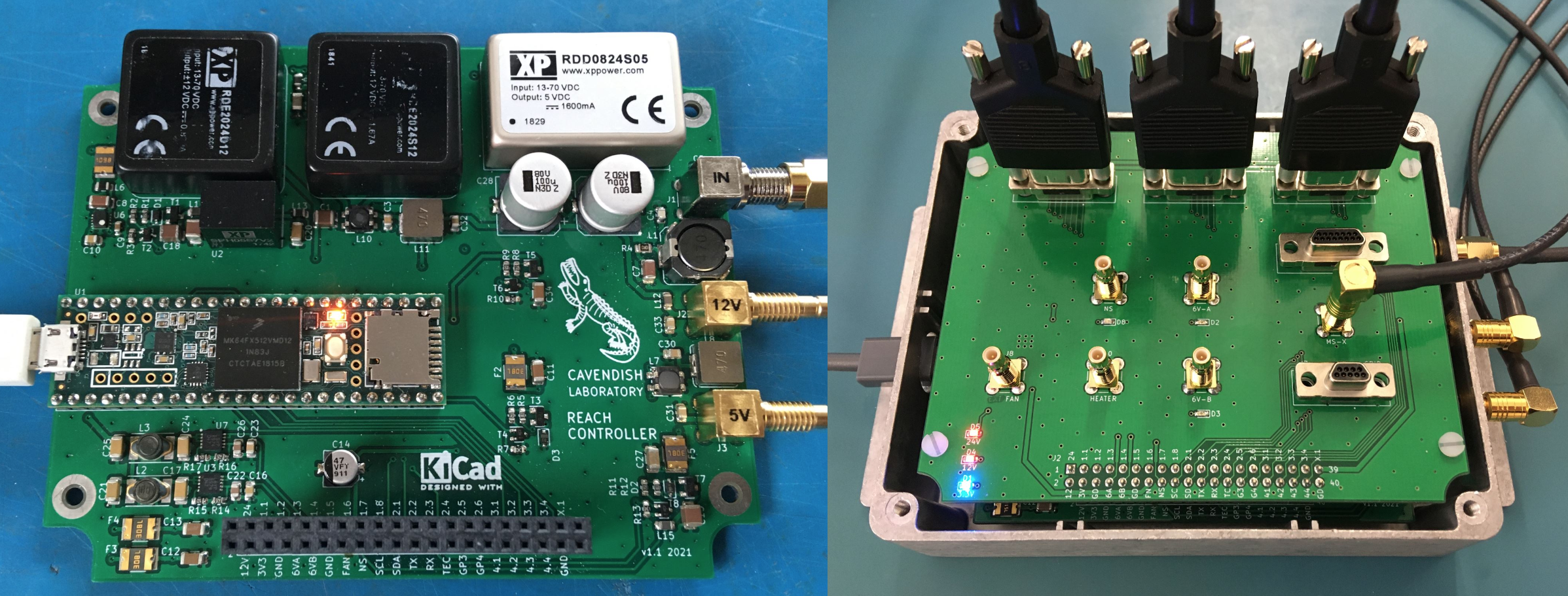}
    \caption{The bottom and top layers of the stacked microcontroller unit. The controller board (left) shows the Teensy microcontroller on the board's centre with power supplies on the right of the board. When housed, the breakout board (right) sits above the controller board as shown and features connection ports, electromagnetic interference filtering and the noise source (28 V) regulator.}
    \label{fig:micro}
\end{figure}

To achieve this functionality in a small form factor, the unit was stacked, with the microcontroller board placed under a custom breakout board which would supply DC power to various components in the box, send control signals to the mechanical switches and provide additional filtering. This arrangement is shown in \cref{fig:micro}. Additional noise reduction measures were applied such as using conductive gaskets placed under bulkhead connectors.

\subsection{Front-end enclosure and thermal considerations}
The REACH front-end enclosure was made using an off-the-shelf $500 \times 500 \times 210$ mm stainless steel IP66 box (Rittal 1007.600) with a hinged door that had an electromagnetic interference gasket placed around the opening to reduce both the impact of self-generated RFI from the box to the antenna as well as external RFI sources feeding into the RF signal chain. The box served two functions, one to be an RFI tight enclosure for all the front-end electronics and the other to help dump heat generated from components inside the box to the outside using a custom heat exchanger with a fan-assisted heatsink.

\begin{figure}
    \begin{subfigure}{0.49\columnwidth}
        \includegraphics[width=\columnwidth]{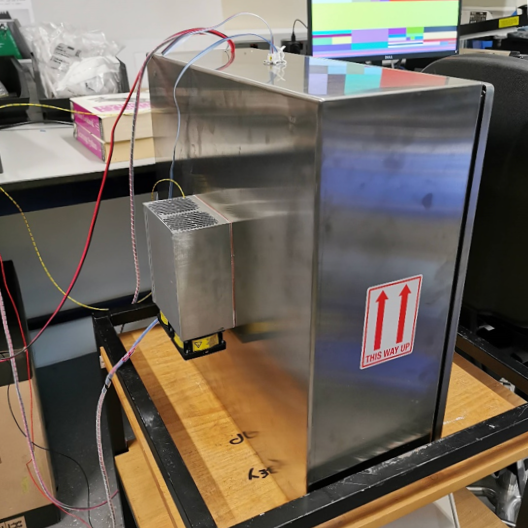}
    \end{subfigure}%
    \hspace*{\fill}  % maximise separation between sub figures
    \begin{subfigure}{0.49\columnwidth}
        \includegraphics[width=\columnwidth]{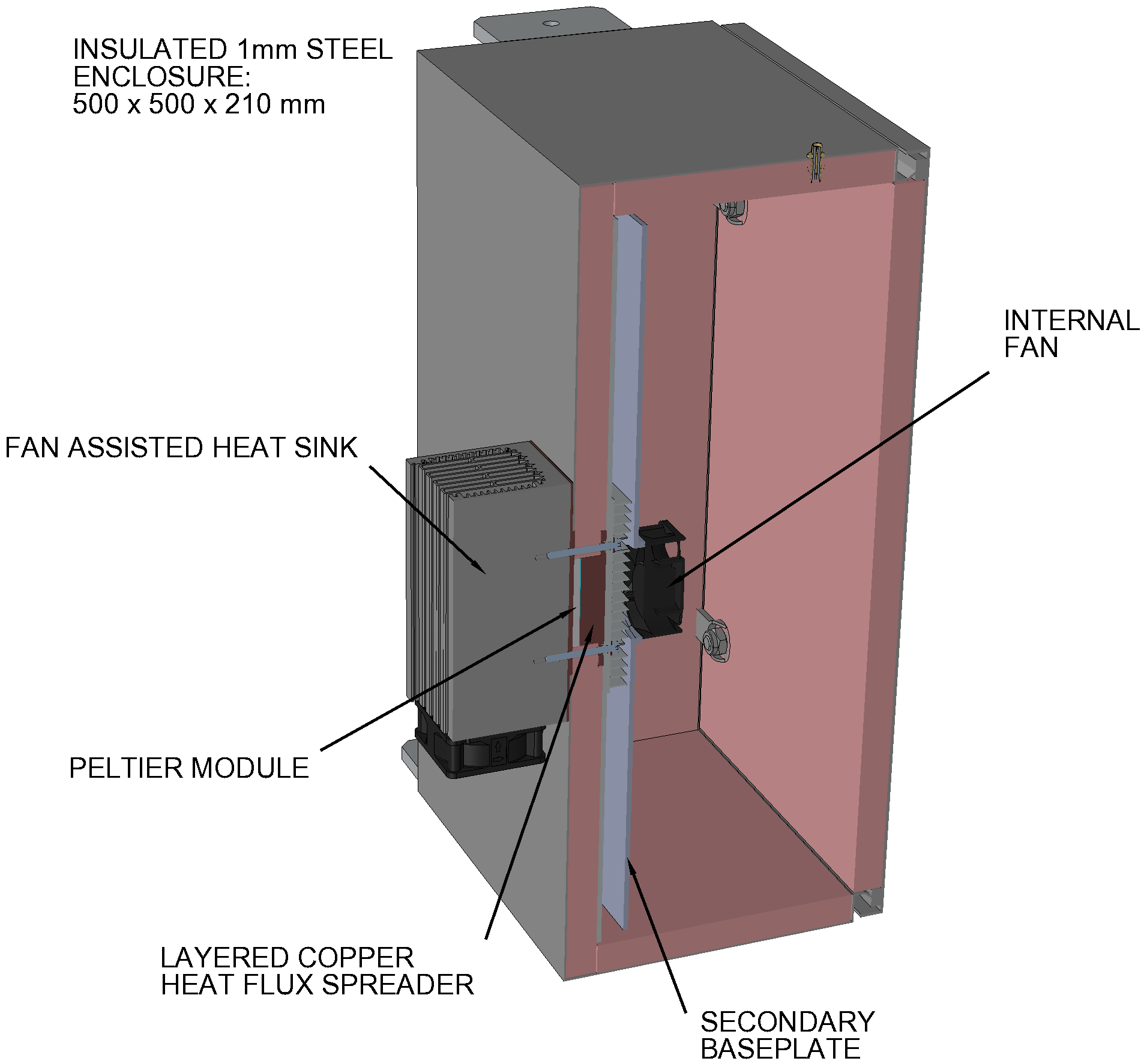}
    \end{subfigure}%
    \caption{The completed front-end thermal enclosure is shown on the left. A 3D-rendered cross section in the same orientation is shown on the right depicting the internal fan, baseplate, Peltier module and heat sink configuration.}
    \label{fig:enclosure}
\end{figure}

To achieve the latter, 20 mm nominal thickness Kooltherm type K5 building insulation panels were attached to all the walls inside the enclosure. The actual closed cell foam thickness was found on measurement to be 18 mm. The receiver components were mounted on a suspended 3 mm baseplate to allow airflow between the baseplate and an internal heat exchanger. The heat exchanger consisted of a 113 W Peltier device coupled to a custom-layered copper thermal stack to spread the heat flux. This stack ended with a larger copper plate attached to the bottom of the box which would help spread the heat to the outer wall and was further cooled using an external heatsink and fan as shown in \cref{fig:enclosure}.

Following experiments done with a 40 W heat source placed in the centre of the plate, an 8 K temperature gradient was observed across the baseplate. To alleviate this, a secondary baseplate and fan were placed between the original baseplate and the receiver’s internal components as rendered in \cref{fig:enclosure}. A negligible temperature variation of 0.125 K was observed across the secondary plate during measurements.

To control the Peltier device, an off-the-shelf thermoelectric cooler (TEC) Proportional-Integral-Derivative (PID) controller (Electron Dynamics TC-M-U-10A) was used. A separate 22 V power supply module was designed to reduce RFI coupling from the very large switch currents produced. This power supply module also automatically powers the external fans when the TEC controller draws more than 6 watts. The TEC controller could be programmed with the temperature set point (typically 30 $^{\circ}$C in the laboratory) as well as the PID parameters. 

\subsection{Back-end observation system}
\begin{figure}
    \centering
    \includegraphics[width=0.75\columnwidth]{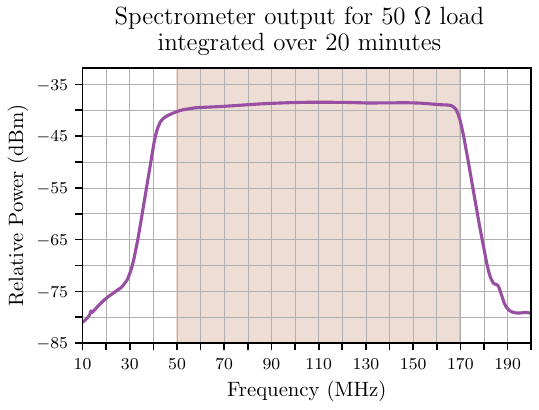}
    \caption{Typical spectrum obtained from integrating on a 50 $\Omega$ load for 20 minutes. A stable power measurement is seen within the REACH observational band (shaded region).}
    \label{fig:spectra}
\end{figure}

The RF front-end which incorporates the in-field calibrator will sit under the antenna as described in \citet{reach}. As previously mentioned both the RF and control signals are relayed back to the back-end system via optical fibres. Following AMP2 in the back-end node, the RF signal is fed into a high-resolution spectrometer based on the SKA1-Low iTPM Field-programmable Gate Array (FPGA) board. The iTPM hosts 16x 14-bit ADC channels (ADI AD9680 chip) and two Xilinx Ultrascale FPGAs. This system digitises the analogue signal at 400 MSPS using 16,384 channels resulting in 12.2 kHz resolution per channel. The iTPM provides a platform for fast development of radio-astronomy digital back-ends. This platform was originally developed in the context of the Aperture Array Verification System for SKA \citep{itpm} and many of the auxiliary functions, such as communication over gigabit Ethernet for monitoring, control and data acquisition, are reused with minimal modifications, while the FPGA firmware has been customised such that each FPGA processes a single digitised RF signal using a floating-point FFT and polyphase filterbank incorporating a total of 229,376 tap coefficients \citep{reach}. Spectra are then typically accumulated over a number of FFT frames corresponding to an integration time of approximately 1 second. These accumulated spectra are then transmitted to the processing server where further accumulation can take place, typically of order minutes. A typical spectrum obtained from a 20-minute integration on a 50 $\Omega$ load is shown in  \cref{fig:spectra}.

The back-end control software is responsible for the automation of all control loops and data taking which is typically initiated by a high-level YAML file. The process typically involves configuring, initialising and controlling various components including the VNA and switches. A typical calibration run includes a number of steps requiring source switching, VNA calibration and measurements as well as spectra accumulation and hardware monitoring. Upon completion of a calibration run, the generated output files can then be transferred off-site through a satellite network link, as described in \citet{reach}.

\begin{figure}
    \includegraphics[width=\columnwidth]{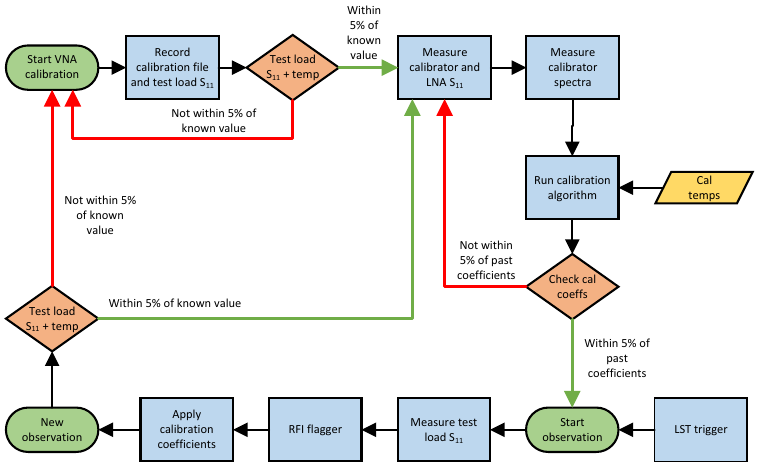}
    \caption{Typical REACH calibration and observation loop including calibration of the VNA and receiver as well as data acquisition, checks and processing such as RFI flagging. Various stages are verified via algorithm before proceeding to the next process.}
    \label{fig:obs}
\end{figure}

\Cref{fig:obs} shows a flow chart for a typical calibration and observation run including calibration of the on-board VNA using S-O-L standards which is verified using the test load before measuring the $S_{11}$ of the sources, antenna and the LNA. This is followed by spectral measurements along with concurrent thermocouple temperature measurements at a cadence of 10 seconds. These data are then used to compute the noise-wave parameters described in \cref{calibration} either on-the-fly or offline. A caveat to the multitude of calibration sources installed is the time taken to cycle all of them through the Dicke switch. A single hour of integration on each calibrator results in 36 hours of total system run time, which extends to 40 hours when including overheads to ensure thermal stability after switch toggling and VNA calibration. 

In a normal observation run, REACH will rely on Dicke switching to observe the sky, ambient load and noise source on regular intervals of 10-30 seconds resulting in much shorter integration periods than typically required for obtaining calibration datasets. It is not yet known how often calibration data will need to be updated to improve estimation of the noise waves which are likely to change if the internal front-end enclosure temperature is changed to a different set point. Initially, we expect half of our data to be used for calibration purposes and the other half for observation of the sky. Furthermore, since the sky signal will be highly isolated (at least 100 dB) from the calibration sources we will rely on to obtain the noise wave parameters, we can treat the process of calibration versus observation as being completely independent of one another.

\subsection{Final deployed system}\label{subcalibration}
\begin{figure}
    \begin{subfigure}{0.5\columnwidth}
        \includegraphics[width=\columnwidth]{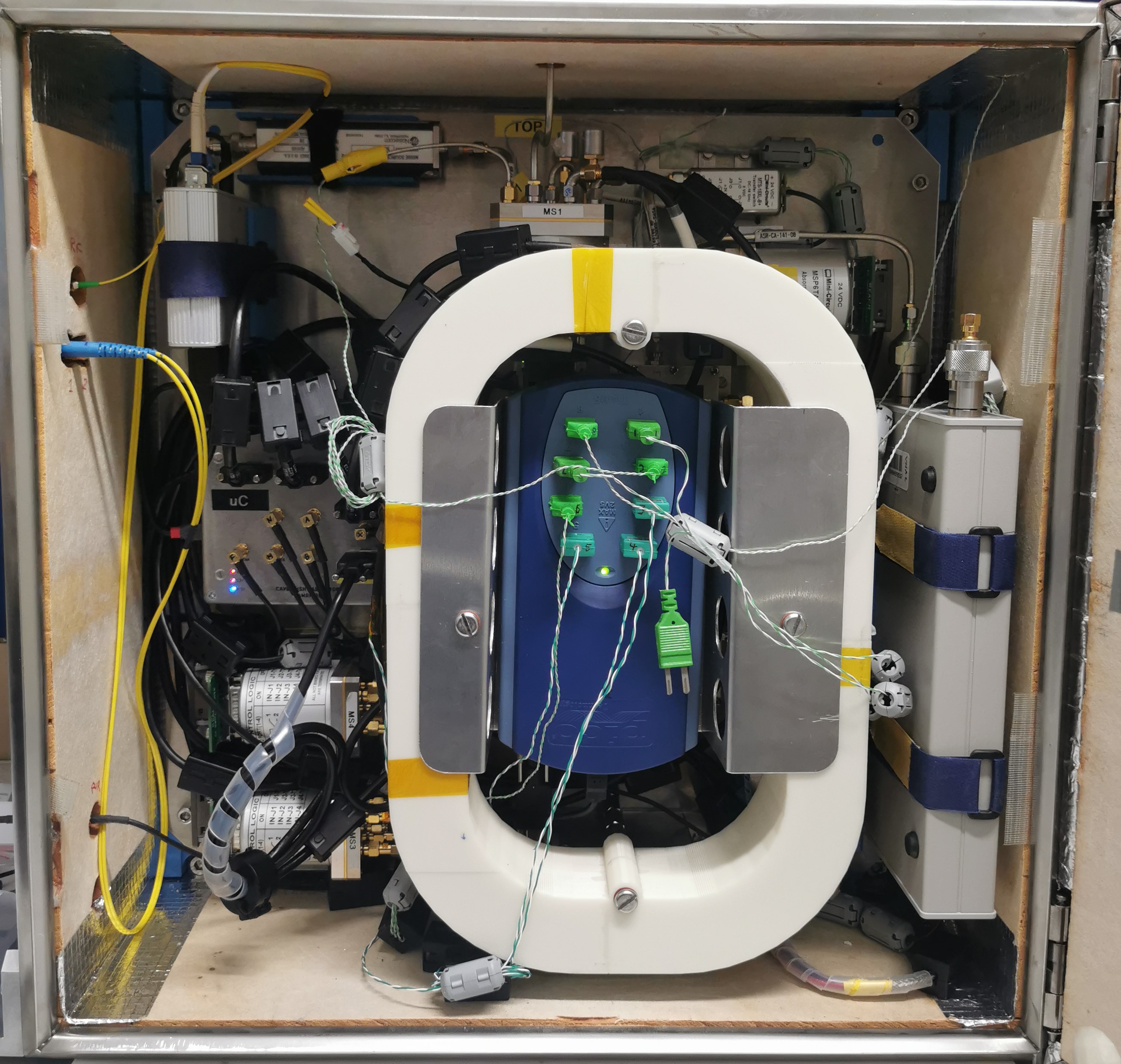}
        \caption{Front-end field unit} \label{fig:deploy_f}
    \end{subfigure}%
    \hspace*{\fill}  % maximise separation between sub figures
    \begin{subfigure}{0.455\columnwidth}
        \includegraphics[width=\columnwidth]{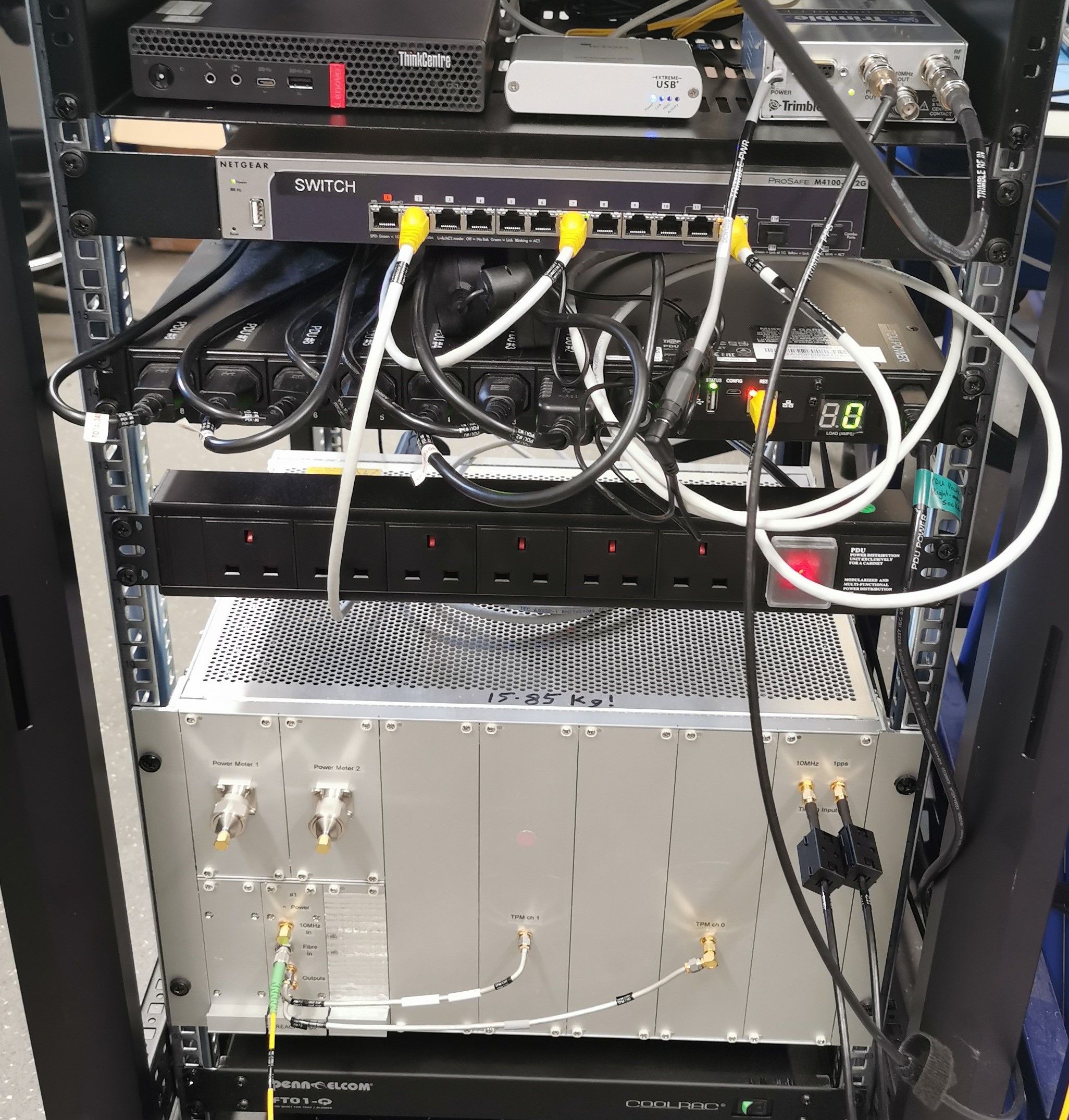}
        \caption{Back-end node rack} \label{fig:deploy_b}
    \end{subfigure}%
    \caption{REACH front-end and back-end deployable hardware. The front end unit (left) shows the compact VNA on the bottom right and the TC-08 module in the centre with green thermocouples. A custom oval-shaped housing for the long cables is seen in white. The various cylinders are the multi-input switches connected to calibration sources. The microcontroller unit can be seen on the middle-left with the RFoF link and diode noise source on the top left corner. The LNA and AMP1 modules housed under the TC-08 cannot be seen. The back-end rack (right) shows the RFoF link to ThinkCentre server, a Trimble Thunderbolt E GPS clock, RJ45 connection switch for communication across back-end devices, power distribution units and a fan cooling rack. The silver installation houses the iTPM spectrometer, AMP2 module, out-of-band noise module and ports for measurement of the back-end power consumption via power meter.} \label{fig:deploy}
\end{figure}

The final field unit and back-end hardware are shown in \cref{fig:deploy}. The field unit (\cref{fig:deploy_f}) houses the various components shown in \cref{fig:overview}. The grey rectangular box on the right-hand side is the TR1300/1 VNA. On the left-hand-side, switches MS3 and MS4 are visible along with the microcontroller unit, USB to Fibre converter, NC346A calibrated noise source as well as the hot load which is simply constructed from a proportional heater placed on a 50 $\Omega$ termination, and a 4-inch RG-405 cable. The main MS1 switch along with MS2 and MTS are also visible towards the top. All RF connections inside the box are made using RG-402 semi-rigid cables for stability.

The central blue box is the TC-08 thermocouple module which measures the temperatures of MS1, MS3, MS4, hot load, 2 m, and 10 m cables as well as the LNA and antenna feeding cable to 0.1 K accuracy. These measurements are required for calibration and are further discussed in \cref{methods}. The white oval ring is a custom-designed 3D printed housing unit for the 2 m and 10 m cables. The LNA, AMP1 and TEC modules are beneath the TC-08 and oval ring and are not visible in this picture. Various RF cable chokes are used in the box to limit radiation from module to module and especially limit any control or power signals from intercepting the RF signal path. This was done through a trial and error process whilst looking at integrated spectra from the system.

In the back-end rack (\cref{fig:deploy_b}), the bottom 6U module is the enclosure housing AMP2 and the iTPM with space available to receive up to two antenna signals. The enclosure is also cooled using an off-the-shelf Peltier-based heat exchanger. Space is available inside the enclosure for up to two power meters which could be used in the field to independently monitor absolute power levels using a USB power meter. This is useful, especially when dealing with in-band RFI on site.

The back-end rack also houses the server (small footprint Lenovo ThinkCentre), USB to fibre converter, and a Trimble GPS unit supplying 10 MHz and 1 PPS signals to the iTPM on the top shelf. There is also an Ethernet-controlled power distribution unit (PDU) and 1G switch for routing data. Fans are placed inside the rack for better heat flow inside the node, which is described in \citet{reach}. 

% ==============================================================================
\section{Additional Corrections}\label{methods}
\subsection{S-parameter corrections} \label{sparams}
Before S-parameter measurements of the calibration sources are made, the VNA itself is calibrated using a standardised Short-Open-Load (SOL) set whose signal path passes through the MS2 switch. The signal path of the calibration sources however includes an additional path length through the MTS switch as shown by the red dashed lines in \cref{fig:overview}. These additional signal paths are also not present during spectral measurements whose paths are represented by the purple dashed lines in \cref{fig:overview}.

To correct for these differing reference planes between MS2-J1 or MS2-J2 and MTS-J2, scattering transfer parameters (T-parameters) are used to numerically de-embed the additional paths \citep{pozar}. For a 2-port network, there is a simple relationship between the S-parameters measured and the T-parameters used, therefore by explicitly measuring the 2-port red paths shown in \cref{fig:overview}, we can determine its T-matrix and move the reference plane accordingly. However, the issue now is that through the switch we still have the purple path which is normally in place. Since we have moved everything to the reference plane defined by MTS-J2, it is easiest to just add this extra path to the amplifier data using the following \citep{pozar}:
\begin{equation}
  \Gamma_{\mathrm{out}} = S_{22} + \frac{S_{12}S_{21}\G{rec}}{1 - S_{11}\G{rec}},
\end{equation}
where $\Gamma_{\mathrm{out}}$ is the corrected LNA reflection coefficient. This forms all the corrections applied to the S-parameters in our system.

\subsection{Temperature calculations} \label{tempModel}
\begin{figure}
    \includegraphics[width=\columnwidth]{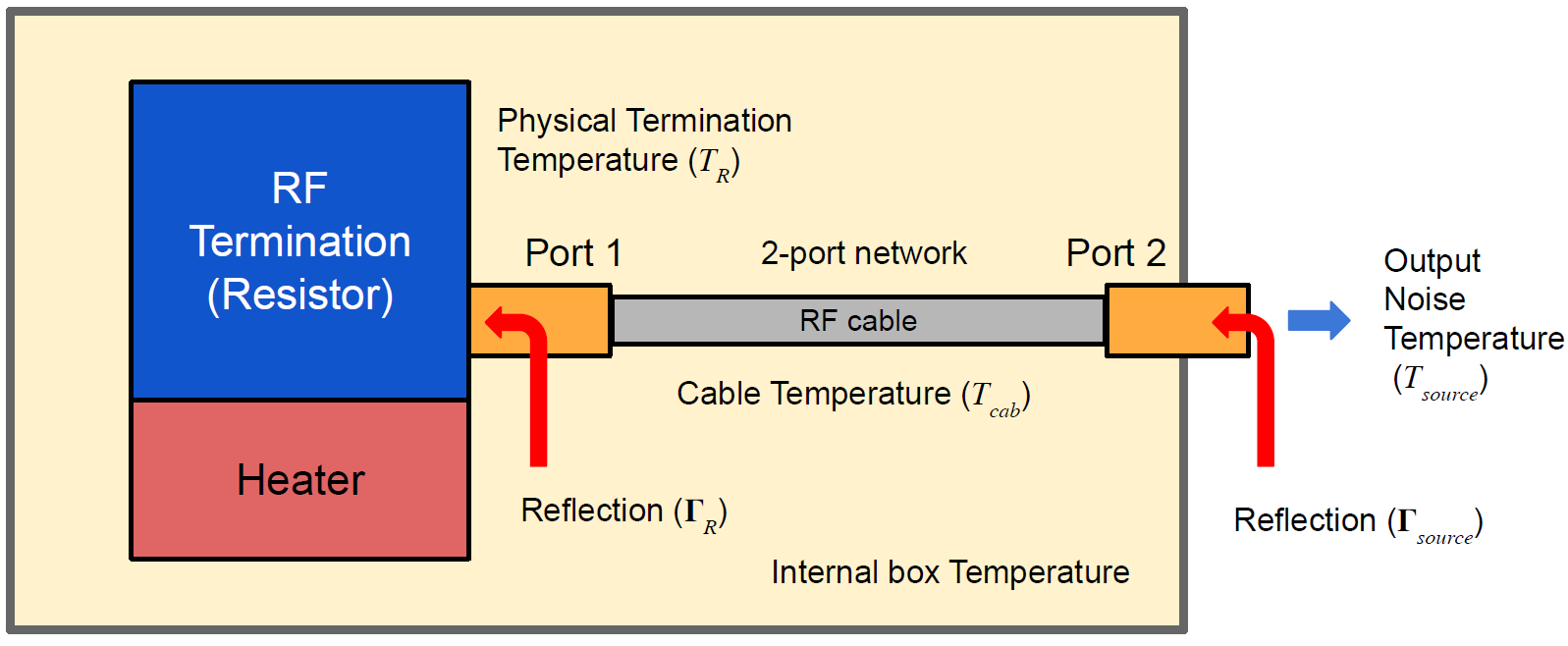}
    \caption{Illustration of the heated load construction for use as a calibration source. The thermally enclosed resistor and powered heater are connected to the input switch via a 4-inch cable as shown in the diagram. These components effectively form a temperature gradient across the calibration device which must be corrected for via the procedure discussed in \cref{tempModel} as $\Tb{R}$, $\Tb{cab}$ and $\Tb{source}$ are not necessarily equal.}
    \label{fig:temp}
\end{figure}

Another key data required by our pipeline is physical temperature. We have normally measured these with a thermocouple at different times during observation and assumed these to be spectrally flat. However, in reality that is not accurate given that multiple sources are comprised of a cable and a source at differing temperatures. This is obviously true for the heated load since the thermal resistor is heated to 370 K whilst the cable attaching it to the system is closer to the internal room temperature (with a temperature slope). \Cref{fig:temp} illustrates this model which is also applicable to all our long cable sources \citep{edgesCal}. For the latter, we have a short cable (2 m) and a long cable (10 m) attached to switches MS3 and MS4, respectively. Four termination resistors are then attached to each switch as shown in \cref{fig:overview} resulting in different impedances with frequency variation at the end of the cable. In the case of the cable calibrators, the termination resistor temperature is heated by the switch (MS3 or MS4) resulting in approximately a 3-degree difference between the cable itself and the load. 

Therefore to correct this behaviour, we must first calculate the realised gain of each source path and apply this to determine the correct source temperature. The realised gain describes the actual gain that a device or system is able to achieve in practice and takes into account factors such as loss and mismatch, given by 
\begin{equation}
  G = \frac{\lvert S_{21} \rvert^2(1- \lvert \G{R} \rvert^2)}{\lvert 1-S_{11}\G{R} \rvert ^2(1- \lvert \G{source} \rvert ^2)},
\label{eq:rgain}
\end{equation}
where $S_{11}$ and $S_{22}$ are the forward S-parameters of the cable. The reflection coefficient is measured both at the resistive load ($\G{R}$) and at the end of the cable ($\G{source}$). The latter is the main S-parameter data discussed in \cref{calibration}. Using this available gain, we can then determine the effective temperature of each source as 
\begin{equation}
  \Tb{source} = G\Tb{Res} + (1 - G)\Tb{cab},
\end{equation}
where $\Tb{R}$ is the temperature of the resistive load itself and $\Tb{cab}$ is the temperature of the cable. Both of these quantities are measured by the thermocouple in normal observations. In our analyses, we found that our calibration was particularly sensitive to this correction at the sub-kelvin scale.

\subsection{Antenna temperature corrections} \label{antModel}
As described in \cref{calibration}, in our calibration pipelines (frequency-by-frequency or Bayesian) we use up to 12 sources to determine the noise waves which are then applied to compute the calibrated temperature of an antenna, whose data has not been used in the calibration process. In the laboratory, this antenna is formed from a 1 m cable and a load attached to it which effectively simulates some frequency variation in impedance that we would normally experience in the field by the real antenna.

By the same principle described in \cref{tempModel}, we are estimating the temperature of this “antenna” at the reference plane (MTS-J2), meaning we have a load at the external ambient temperature and a cable (some of which is inside the receiver box) at another temperature. In the laboratory, the overall cable can be described to have a temperature that is approximately 3 degrees higher than the load temperature. As such we expect the antenna data to exhibit features on the scale of the cable length as with the cable-based calibration sources. To approximately correct this and estimate the antenna temperature at the load end, we must compute the available gain shown in \cref{eq:rgain} which relies on the antenna cable and reflection coefficient measured both at the load end and at the cable end. Using this, we can then rearrange \cref{eq:rgain} to get the following temperature
\begin{equation}
  \Tb{final} = \frac{1}{G}(\Tb{ant}+(G-1)\Tb{cab}).
\end{equation}

If calibration and data correction is performed in the right way, we would expect this temperature ($\Tb{final}$) not to exhibit spectral features. To get a rough handle on $\Tb{cab}$ for this system, we can use the following approximation
\begin{equation}
  \Tb{cab} = \frac{1}{\Lb{tot}} (\Tb{int}\Lb{int} + \Tb{ext}\Lb{ext}),
\end{equation}
where $\Tb{int}$ and $\Tb{ext}$ represent the internal and external cable temperatures, whilst $\Lb{int}$ and $\Lb{ext}$ represent the internal and external cable length, respectively. Here $\Lb{tot}$ is the total cable length. In our case, $\Lb{ext}$ is 100 cm and $\Lb{int}$ is 28 cm. 

In the final field deployment, these corrections are made even more complex due to the response of the balun. Assuming the radiation efficiency of the antenna is 1 (i.e. $\Tb{ant} = \Tb{sky}$), then there will be two RF networks in between the antenna and our well-defined reference plane, namely the balun and the antenna feeding cable. Both of these will be cooler than the hot sky (at the REACH frequencies) and will have the potential to impose frequency structure on the calibrated antenna temperature unless corrections are made.

% ==============================================================================
\section{Calibration data analysis}\label{results}
To assess the performance of the final deployable system, a mock antenna constructed from a 1 m cable attached to an 89 $\Omega$ load at the receiver input was calibrated. With the TEC set to stabilise the internal receiver temperature at 30 $^{\circ}$C, fifteen separate training datasets were collected with 20-minute integrations for spectral measurements and the data was masked to the 50--130 MHz band. Following this, the below procedure was applied to the data;
\begin{enumerate}
    \item Corrections described in \cref{sparams} were applied to all the S-parameter data to correct the reference plane.
    \item Temperature models detailed in \cref{tempModel} were applied to the 2 m and 10 m cable sources as well as the heated load, whilst other sources were assumed to have a flat temperature. The physical temperatures used to form these models were obtained from the TC-08 thermocouple averaged over the data collection period. 
    \item The $\mathbf{X}$-terms were computed using the S-parameter and spectral data.
    \item As an optional step, these $\mathbf{X}$-terms were smoothed, using a smoothing spline factor of 0.999999999 (a factor of 1 being no smoothing). Whilst this step is not critical, it does reduce noise on the measurement data whilst leaving any spectral features that exist in that data.
    \item A least squares approach was taken to solve \cref{eqn:linearmodel} separately for each of the fifteen datasets using a Moore-Penrose pseudoinverse to determine the noise wave parameters on a frequency-by-frequency basis.
    \item Artefacts in four bands across the REACH dipole passband and were excised (replaced with NaN's) from the noise wave parameter calculated in the previous step. This corresponded to a total of 6.2 MHz of data. The artefacts are caused by the $\mathbf{X}$-terms used in the calibration equation going to zero and resulting in a poor fit when compared across all calibrators.
    \item The fifteen noise wave parameter sets were then averaged
    \item These noise wave parameters are then used along with the $\mathbf{X}$-terms to compute the final calibrated temperatures of sources.
    \item For the antenna temperature, the correction discussed in \cref{antModel} can be applied to improve the final result.
\end{enumerate}

\begin{figure}
    \includegraphics[width=\columnwidth]{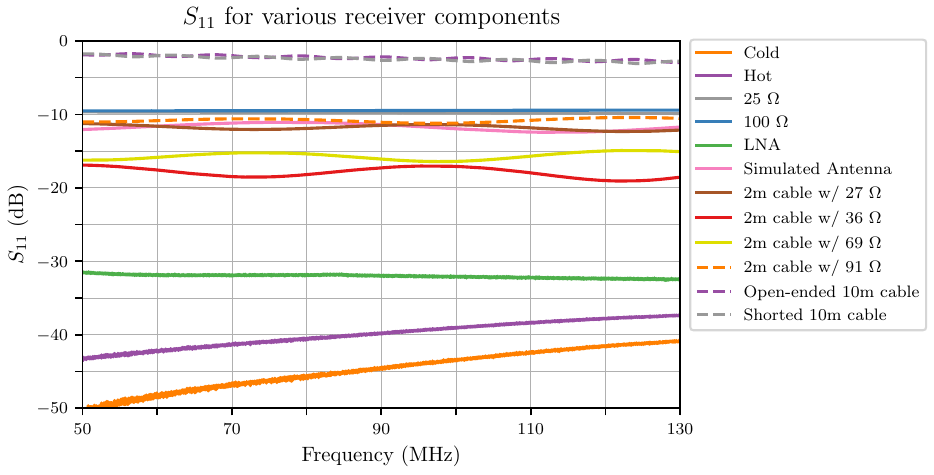}
    \caption{$S_{11}$'s for the various receiver components including the LNA and simulated antenna from a single data file of the fifteen files collected. The entire extent of the plot lies within the REACH observational band.}
    \label{fig:s11plot}
\end{figure}

\begin{figure}
    \includegraphics[width=\columnwidth]{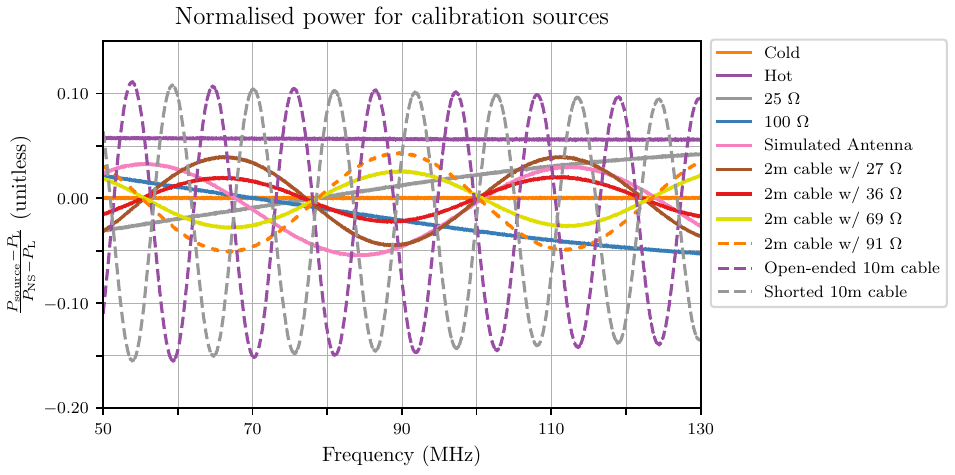}
    \caption{PSD quotient for the calibration sources and simulated antenna from a single data file of the fifteen files collected. The entire extent of the plot lies within the REACH observational band.}
    \label{fig:qplot}
\end{figure}

\begin{figure}
    \centering
    \includegraphics[width=0.85\columnwidth]{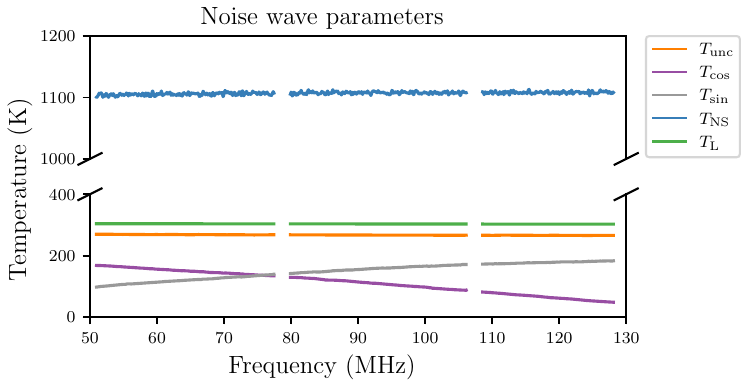}
    \caption{Noise wave parameters derived from one of the fifteen datasets using a frequency-by-frequency least squares fit. Anomalous bands have been excised. The entire extent of the plot lies within the REACH observational band. We recover the expected value for $T_{\mathrm{L}}$ as the noise temperature of the internal load ($\sim300$ K) and $T_{\mathrm{NS}}$ as the excess noise temperature of the internal noise source ($\sim1100$ K).
    }
    \label{fig:nwplot}
\end{figure}

Example plots of S-parameters and PSD quotients which make up the bulk of the data input to the pipeline are shown in \cref{fig:s11plot} and \cref{fig:qplot} respectively. This data is used to compute the constants in \cref{eqn:linearmodel}. Using this data along with the measured temperatures of the sources, we can compute the noise wave parameters using a least squares fit, relying only on 10 of the 12 calibrators (2 m + 10 $\Omega$ and 2 m + 250 $\Omega$ excluded). We have excluded the data from these two devices due to inconsistencies found with the other calibration sources. We acknowledge that this may raise concerns about the non-objectivity of selecting measurements, but emphasise the goal of obtaining the most accurate and reliable calibration solution possible and recognise that this may warrant further investigation in future work. We remain confident that our decision to exclude these data points were justified based on our scientific goals and the quality of the data. \Cref{fig:nwplot} shows the raw noise wave parameters using only a 20-minute integrated dataset.

\begin{figure}
    \centering
    \includegraphics[width=.90\columnwidth]{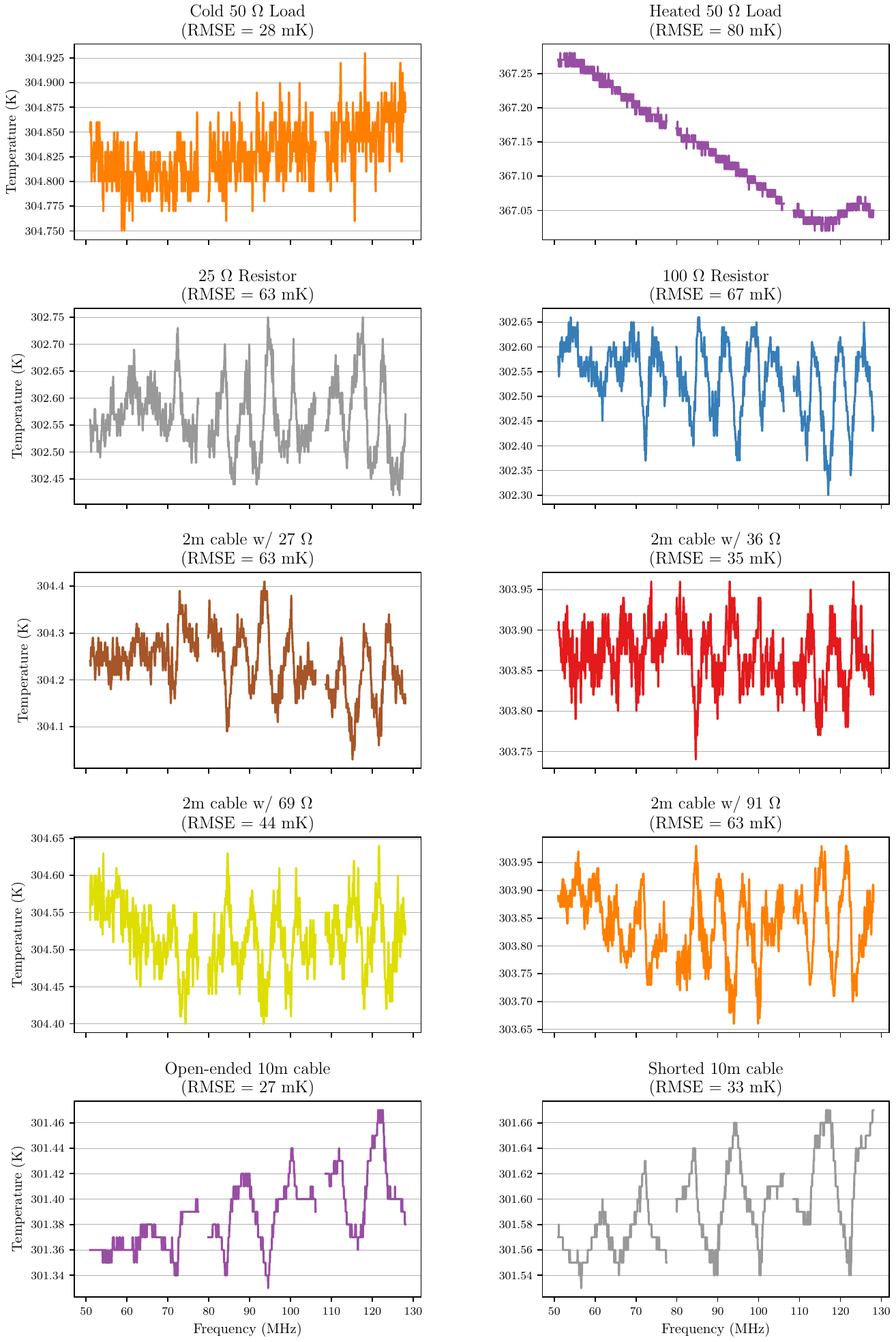}
    \caption{Calibrated temperatures of the various calibration sources using the noise wave parameter values averaged over the fifteen datasets taken. Anomalous data has been excised. The extent of each subplot lies within the REACH observational band.}
    \label{fig:tcal_sources}
\end{figure}

A set of noise wave parameters was calculated for each of the fifteen datasets using the method outlined above which was applied to the calibration sources within each respective dataset to produce fifteen calibrated temperature solutions for each calibrator. The fifteen solutions for each calibrator were then averaged produce a final calibrated temperature for each source as shown in \cref{fig:tcal_sources}.

\begin{figure}
    \includegraphics[width=\columnwidth]{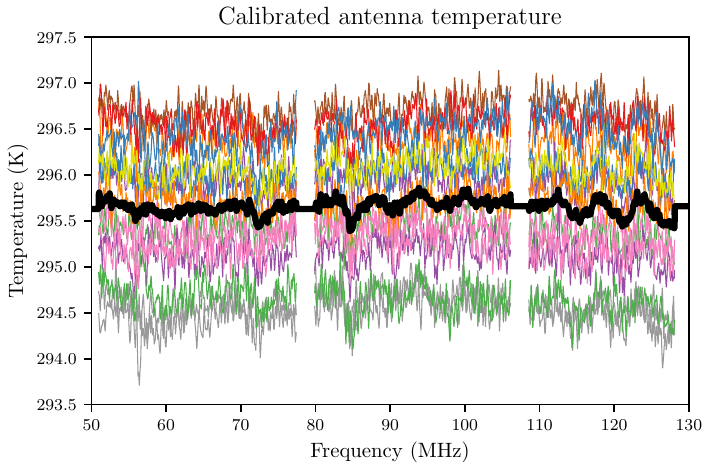}
    \caption{Calibration solutions from each of the fifteen datasets applied to the simulated antenna are shown in various colours. An averaged solution from the fifteen datasets applied to the simulated antenna is shown in black and has an RMSE of 80 mK. The Gaussian structure of the noise converging on an approximately-flat averaged temperature over frequency is seen as encouraging. Anomalous data has been excised and the extent of the plot is within the REACH observational band.}
    \label{fig:tcal_ant}
\end{figure}

The same process was used to determine the final “antenna” temperature but an extra step was taken to correct for the antenna cable as described in \cref{antModel}. \Cref{fig:tcal_ant} shows the calibrated antenna temperature computed for each dataset as well as an average of the fifteen solutions shown in black where an RMSE of 80 mK was obtained. It should be noted that over the period in which all the data was collected (approximately two weeks), the laboratory environmental temperature was observed to vary by up to 3.5 degrees over night and day. 

It has been noted that the results of \cref{fig:tcal_ant} exhibit non-Gaussian structure on the frequency scale of about 5 MHz. These features likely arise due to interactions between the antenna and receiver as well as impedance mismatch contributions from the calibration loads and other environmental factors which will be the focus of future investigations. Additionally, it has been pointed out that the level of noise for our mock antenna is comparable to that of \citet{edgesNature} (figure 1b) which may present difficulties in challenging the EDGES results upon application of our methods to a finalised antenna. We however advise that the simulated value of 300 K corresponds to the impedance-matched condition of our custom antenna mimicking the performance of real-world antennas operating under sky-noise conditions. Furthermore, in practice, it is expected that averaging across more datasets will reduce the RMSE in the final calibrated temperature and this is encouraging for the REACH system being deployed. Methods are currently being investigated for how to update the calibration noise wave parameters from day to day and get accurate results over certain observation periods. We also acknowledge the fine frequency structure present in \cref{fig:tcal_sources} and \cref{fig:tcal_ant} which may be attributed to a number of aspects including noise or reflections generated at the fibre-optic conversion points. During the design and testing of the radio receiver, measurements were conducted to ensure a minimal impact from the fibre-optic conversion on the signal chain, however further analysis may be required to fully understand these effects in the deployment environment. These will form the basis of a follow-up paper which will aim to improve upon the approach taken here.

% ==============================================================================
\section{Conclusion and further work}\label{conclusions}
In this work, we have discussed the receiver design approaches for the REACH experiment targeting the cosmic 21-cm signature from the Cosmic Dawn. In order to achieve a sufficient level of calibration with a focus on an in-field calibration using minimal laboratory-based calibration data, multiple practical techniques have been incorporated into the radiometer architecture including a front-end housing up to twelve calibration sources connected to low-loss mechanical switches for strategic sampling of the frequency dependent noise waves. Additionally, three custom made amplifiers were constructed to achieve key benchmarks such as an LNA input $S_{11}$ lower than 30 dB to reduce the impact of noise waves on cosmological data. Furthermore, a custom microcontroller unit for the radiometer power management was carefully designed for functionality while maintaining a small form factor by stacking the controller and breakout boards. Additional custom parts were incorporated into the assembly for maximal functionality, such as the thermal stack for heat management which realises an 8 kelvin temperature gradient across the $450 \times 470$ mm baseplate as well as 3D printed parts for housing and RFI mitigation.

In order to facilitate a successful detection of the cosmic signature a calibration methodology incorporating the Dicke switching technique was used for measurements of the PSDs which, along with reflection coefficient and temperature measurements, allow us to derive noise wave parameters that characterise the frequency response of our instrument. Included in the methodology are corrections to the data such as the de-bedding of extra signal paths between the MS2 and MTS switches. Along with this were temperature measurement corrections to incorporate the spectral variation in temperature due to the 3 degree difference between the components of various calibration sources and the model antenna used in our analysis.

These techniques have been applied to a calibration run incorporating ten calibration standards calibrated against a model antenna with $S_{11}$ similar to that of the deployed REACH antenna. Fifteen 20-minute integration runs were performed to compute the noise wave parameters after application of a smoothing spline and excision of troublesome data within the full bandwidth, which was then cut to 50–130 MHz and solved through a least squares approach. The resulting fifteen noise wave parameter sets were then averaged to obtain the final calibration solution. This calibration solution achieves an average RMSE of 59 mK when applied to the various calibration sources and an 80 mK RMSE for the model antenna. We also report a 30 mK RMSE for our calibration solution applied to the sources attached to 10 metre cables. This calibration level is comparable to that of the EDGES installation \citep{edgesCal}.

Challenges observed during our experiments were the difficulties maintaining environmental stability over time as the many sources incorporated in our calibration technique increased the time for data collection. With the laboratory environment temperature varying by up to 3.5 K, we expect these effects to be amplified when deployed to South Africa. We also note room for possible improvement to our system in future experiments such as the use of fewer switches to combat the complicated modelling of the cables included in the system or the development of a better LNA to lower the impact of noise waves while maintaining good input matching. A better VNA calibration may be achieved through incorporation of better SOL standards which would directly improve the overall calibration of the system. Longer datasets for further noise wave parameter averaging may be incorporated for better performance as well. Introducing an additional simulated antenna with higher noise such as through a heated resistor or electronic noise source may also offer valuable insights into the receiver’s performance under more extreme conditions.

Future work regarding these techniques would be to evaluate the effect of the REACH antenna deployed in South Africa which includes additional cables and a balun not included in our models. Bayesian RFI mitigation techniques could also potentially be incorporated into the REACH pipeline to manage any remaining instrumental sources of RFI \citep{samsPaper}. We also propose exploration of further machine learning techniques to improve the derivation of the calibration parameters and overall characterisation of the system which will be addressed in future works.

\bmhead{Acknowledgements}
WJH was supported by a Royal Society University Research Fellowship. EdLA was supported by a STFC Ernest Rutherford Fellowship. The REACH collaboration acknowledges the Kavli Institute for Cosmology in Cambridge, Stellenbosch University, the National Research Foundation of South Africa and the Cambridge-Africa ALBORADA Research Fund for their financial support of the project.

\bmhead{Data availability}
The data that support the findings of this study are available from the corresponding author upon reasonable request.

\section*{Statements and Declarations}
\bmhead{Competing Interests}
The authors declare no conflicts of interest.

\bibliography{references}

\end{document}